\documentclass[12pt]{iopart}
\usepackage{graphicx}
\usepackage{dcolumn}
\usepackage{bm}
\usepackage{epstopdf}
\usepackage[dvipsnames]{xcolor}
\usepackage{soul}
\usepackage[utf8]{inputenc}
\usepackage[T1]{fontenc}
\usepackage{indentfirst}
\usepackage{enumitem}
\usepackage{xspace}
\usepackage{hyperref}
\hypersetup{breaklinks=true,colorlinks=true,linkcolor=blue,citecolor=blue,filecolor=magenta,urlcolor=cyan}
\usepackage{datetime}

\usepackage{amssymb}
\usepackage{color}
\usepackage{array,multirow}

\newcommand{\newblock}{}
\newcommand{\notag}{\nonumber}

\newcommand{\tfrac}[2]{{\textstyle\frac{#1}{#2}}}
\newcommand{\text}[1]{\mbox{\scriptsize{#1}}}

\newcommand{\spc}{{\ }}
\newcommand{\pr}[1]{{\sc{\lowercase{#1}}}}

\newcommand{\gras}[1]{\boldsymbol{#1}}

\newcommand{\codeversion}        {3.06h}
\newcommand{\execversion}        {306h}
\newcommand{\sizeversion}        {7}
\newcommand{\modulesversion}     {35}
\newcommand{\hfbthoversion}      {201}
\newcommand{\interfaceversion}   {5}
\newcommand{\functionalversion}  {4}
\newcommand{\mpiioversion}       {6}
\newcommand{\mpimanagerversion}  {5}
\newcommand{\shellversion}       {5}
\newcommand{\SLsizversion}       {4}
\newcommand{\fissionversion}     {9}
\newcommand{\pairsversion}       {2}
\newcommand{\pnpversion}         {8}
\newcommand{\fitsversion}        {16}
\newcommand{\lipkinversion}      {31}
\newcommand{\tgradversion}       {18}
\newcommand{\wignerversion}      {6}

\newcounter{leteq}

\newenvironment{eqnalpha}{\setcounter{leteq}{1}

\begin{eqnarray}}{\end{eqnarray}%
}

\newenvironment{eqnalphalabel}[1]{\setcounter{leteq}{1}
\raisebox{0cm}[0cm][0cm]{\begin{minipage}{1cm}%
\begin{eqnarray}\label{#1}&&\nonumber\end{eqnarray}\end{minipage}}

\begin{eqnarray}}{\end{eqnarray}%
}

\newcommand{\bnl}{\begin{eqnalpha}}
\newcommand{\enl}{\end{eqnalpha}}
\newcommand{\bnll}[1]{\begin{eqnalphalabel}{#1}}
\newcommand{\enll}{\end{eqnalphalabel}}

\newcommand{\newton}[2]{\left(\raisebox{-1ex}{$\stackrel{\textstyle{#1}}{#2}$}\right)}

\newcommand{\keyw}{{\bf Keyword:}}
\newcommand{\key}[1]{\vspace{1ex}\noindent\keyw{\spc}{\tk{#1}}
                         \newline\phantom{\keyw{\spc}{\tk{#1}}}{\spc}}

\newcommand{\keyspace}{\vspace{2ex}}
\renewcommand{\key}[1]{\vspace{1ex}\noindent\color{red}\keyw{\spc}{\tk{#1}}\color{black}
                         \newline\phantom{\keyw{\spc}{\tk{#1}}}{\spc}}

\newcommand{\be}{\begin{equation}}
\newcommand{\ee}{\end{equation}}
\newcommand{\ba}{\begin{array}}
\newcommand{\ea}{\end{array}}
\newcommand{\bn}{\begin{eqnarray}}
\newcommand{\en}{\end{eqnarray}}
\newcommand{\bc}{\begin{center}}
\newcommand{\ec}{\end{center}}

\newcommand{\thalf}{{\textstyle{\frac{1}{2}}}}

\renewcommand{\rmd}{{\rm d}}

\newcommand{\ti}[1]{#1\index{#1}}

\newcommand{\tv}[1]{{\tt{#1}}\index{#1}}
\newcommand{\tk}[1]{{\tt{#1}}\index{#1}}

\newcommand{\tf}[1]{{\tt{#1}}\index{#1}}

\begin{document}

\title[Solution of universal nonrelativistic nuclear DFT: program \pr{hfodd} {(v\codeversion)}]
{Solution of universal nonrelativistic nuclear DFT equations in
the Cartesian deformed harmonic-oscillator basis.
(IX) \pr{hfodd} {(v\codeversion)}: a new version of the
program.
}

\author{
J.~Dobaczewski,$^{1-4}$
P.~B\k{a}czyk,$^{3}$
P.~Becker,$^{1}$
M.~Bender,$^{5}$
K.~Bennaceur,$^{5}$
J.~Bonnard,$^{1}$
Y.~Gao,$^{2}$
A.~Idini,$^{6}$
M.~Konieczka,$^{3}$
M.~Kortelainen,$^{2,4}$
L.~Pr{\'o}chniak,$^{7}$
A.M.~Romero,$^{1,8}$
W.~Satu{\l}a,$^{3,4}$
Y.~Shi,$^{2,9}$
T.R.~Werner$^{3}$ and
L.F.~Yu$^{2}$
}

\address{$^1$ Department of Physics, University of York, Heslington, York YO10 5DD, United Kingdom}
\address{$^2$ Department of Physics, P.O. Box 35 (YFL), FI-40014 University of Jyv\"askyl\"a, Finland}
\address{$^3$ Institute of Theoretical Physics, Faculty of Physics, University of Warsaw,
              ul. Pasteura 5, PL-02093 Warsaw, Poland}
\address{$^4$ Helsinki Institute of Physics, P.O. Box 64, FI-00014 University of Helsinki, Finland}
\address{$^5$ Univ Lyon, Universit\'e Claude Bernard Lyon 1, CNRS, IP2I, UMR 5822,
              4 rue E. Fermi, F-69622 Villeurbanne Cedex, France}
\address{$^6$ Division of Mathematical Physics, Department of Physics, LTH, Lund University,
              P.O. Box 118, S-22100 Lund, Sweden}
\address{$^7$ Heavy Ion Laboratory, University of Warsaw, PL-02093 Warsaw, Poland}
\address{$^8$ Department of Physics and Astronomy, University of North Carolina,
              Chapel Hill, North Carolina, 27516-3255, USA}
\address{$^9$ Department of Physics, Harbin Institute of Technology, Harbin 150001, China}

\begin{abstract}
We describe the new version {(v\codeversion)} of the code HFODD that solves
the universal nonrelativistic nuclear DFT Hartree-Fock or
Hartree-Fock-Bogolyubov problem by using the Cartesian deformed
harmonic-oscillator basis. In the new version, we implemented the
following new features: (i) zero-range three- and four-body central
terms, (ii) zero-range three-body gradient terms, (iii) zero-range
tensor terms, (iv) zero-range isospin-breaking terms, (v)
finite-range higher-order regularized terms, (vi) finite-range
separable terms, (vii) zero-range two-body pairing terms, (viii)
multi-quasiparticle blocking, (ix) Pfaffian overlaps, (x)
particle-number and parity symmetry restoration, (xi) axialization,
(xii)  Wigner functions, (xiii) choice of the harmonic-oscillator
basis, (xiv) fixed Omega partitions, (xv) consistency formula between
energy and fields, and we corrected several errors of the previous
versions.
\end{abstract}

\submitto{\JPG}

\maketitle

\section{Introduction}
\label{sec:introduction}

The method of solving the Hartree-Fock (HF) equations in the Cartesian harmonic
oscillator (HO) basis was presented in Ref.~\cite{(Dob97b)}.
Seven versions of the code \pr{hfodd} were previously published in seven independent
publications:
(v1.60r)~\cite{(Dob97c)},(v1.75r)~\cite{(Dob00d)}, (v2.08i)~\cite{(Dob04d)},
(v2.08k)~\cite{(Dob05h)}, (v2.40h)~\cite{(Dob09g)}, (v2.49t)~\cite{(Sch12c)}, and (v2.73y)~\cite{(Sch17d)}.
Version (v2.08i)~\cite{(Dob04d)} introduced solutions of the Hartree-Fock-Bogolyubov (HFB) equations.
Below we refer to these publications by using roman capitals II--VIII.
We also acknowledge earlier applications of the Cartesian deformed harmonic-oscillator basis
to the solution of the nuclear self-consistent problem \cite{(Gog75b),(Koe89)}.
The User's Guide for version (v2.40h) is available in Ref.~\cite{(Dob09h)}
and the User's Guide for the present version {(v\codeversion)}
will be published in Ref.~\cite{(Dob21a)}. The full distribution of
the version
{(v\codeversion)} of the code \pr{hfodd} can be found in the Supplemental Material.
The code home page is at
\href{http://www.fuw.edu.pl/~dobaczew/hfodd/hfodd.html}
     {http://www.fuw.edu.pl/\~{ }dobaczew/hfodd/hfodd.html},
and the code repository is at
\href{https://webfiles.york.ac.uk/HFODD/}
     {https://webfiles.york.ac.uk/HFODD/}.
The repository is meant to serve as the first point of contact for
users wishing to run the code. It contains the full downloadable
distribution of the version {(v\codeversion)} of the code \pr{hfodd}
along with numerous examples of the input data files accompanied with
the corresponding output files. In the future, the repository will be
dynamically upgraded; it will contain future distributions of the
code along with any bugfixes implemented before the next version of
\pr{hfodd} will have been published in a journal. It will also
contain descriptions of new features and examples of new input data
files.

The present guide is a long write-up of the new version
{(v\codeversion)} of the code \pr{hfodd}. This extended version
supersedes all previous versions of the program. It features a number
of new implementations listed in Section~\ref{sec:modifications}. In
the serial mode, the new version {(v\codeversion)} of the code
\pr{hfodd} remains fully compatible with all previous versions. One
should note, however, that in the new version {(v\codeversion)},
features of the parallel mode were not thoroughly tested, and the new
developments are not recommended for use in the parallel mode. In the
same way, options related to temperature or fission were not in the
main focus of the present developments and should be considered
fragile. Otherwise, information provided in previous publications
\cite{(Dob97b)}--\cite{(Sch17d)} remains valid, unless explicitly mentioned in the present long
write-up.

The user must have access to various BLAS, LAPACK, and LINPACK
subroutines, which should be either pre-installed on a given system
or downloaded from the Netlib Repository at the University of
Tennessee, Knoxville: \href{http://www.netlib.org/}{http://www.netlib.org/}. Otherwise,
generic versions of subroutines are also included in the \pr{hfodd} distribution,
available in the Supplemental Material and from the code repository at
\href{https://webfiles.york.ac.uk/HFODD/}
     {https://webfiles.york.ac.uk/HFODD/},
and can be compiled along with the main program and its modules.

Version {(v\codeversion)} of the code \pr{hfodd} is free software:
anyone can redistribute it and/or modify it under the terms of the GNU
General Public License as published by the Free Software Foundation,
either version 3 of the License, or any later
version. Code \pr{hfodd} is distributed in the hope that it will be useful, but
{\em without any warranty}; without even the {\em implied warranty of
merchantability or fitness} for a particular purpose. See the
GNU General Public License
\href{http://www.gnu.org/licenses/}
     {http://www.gnu.org/licenses/},
for more details. The authors would gladly receive any communication
regarding the code, however, no dedicated workforce is available for
providing user support of any kind.

In Section \ref{sec:modifications}, we review the modifications
introduced in the version {(v\codeversion)} of the code \pr{hfodd}.
Section \ref{sec:input_file} lists all additional new input keywords
and data values, introduced or modified in version {(v\codeversion)}.
The rules of building the input data file were defined in
Section~II-3~\cite{(Dob97c)} and in the serial mode of version
{(v\codeversion)} they remain exactly the same. These rules specify
the generic structure of the input data file, irrespective of which
specific keywords are used. In particular, the keywords can be read
in any order and, unless explicitly stated in their description, they
are independent of one another. In every new version of the code
\pr{hfodd}, the list of keywords grows and covers new
implementations, but they always abide by the same rules specified in
Section~II-3~\cite{(Dob97c)}. With the new keywords introduced in the present
version {(v\codeversion)} of the code \pr{hfodd}, the list of available
keywords already contains 311 items and their descriptions are scattered
over nine different publications. The new User's Guide \cite{(Dob21a)}
will contain comprehensive coverage of the complete information.

\section{Modifications introduced in version {(v\codeversion)}}
\label{sec:modifications}

\subsection{Zero-range three- and four-body central terms}
\label{subsec:3+4}

In his seminal article~\cite{(Sky59b)}, T.~H.~R.~Skyrme suggested to complement
an effective two-body interaction with a contact three-body term. He also
underlined the fact that such a three-body interaction, averaged over one
of the particles, gives a contribution to the two-body contact term
proportional to the local scalar density $\rho_0$.
This observation motivated the use
of a two-body contact density-dependent two-body term by Vautherin and
Brink~\cite{(Vau72c)}. Their interaction was later extended to include the
possibility to have different weights for its spin-direct and spin-exchange
parts and, possibly, a nonlinear dependence on the density
with a power $\alpha$. It
is usually written as
\begin{equation}
\hat v_3(i,j)=
\tfrac{1}{6}\,t_3 \left(1+x_3 \hat{P}_{ij}^\sigma \right)
 \, \rho_0^\alpha(\mathbf R_{ij}) \,
\delta(\mathbf r_{ij}),
\end{equation}
where $\mathbf R_{ij}$ and $\mathbf r_{ij}$ are the center-of-mass and
relative positions of the interacting particles. This general density-dependent
two-body interaction cannot be related with an underlying three-body force
(even in the case where $x_3$ and $\alpha$ are set to 1) but provides
a phenomenological representation of many-body effects~\cite{(Vau72c)}.
We recall that this term does not completely obey the Pauli
exclusion principle and it generates self-interaction terms.

It was pointed out that the latter features might prevent
unambiguous implementation of the multi-reference (MR) extensions of nuclear energy
density  functionals~\cite{(Ang01),(Rob07d),(Dob07),(Lac09),(Ben09),(Sat14g)}.
In addition, non-integer values of  $\alpha$ can lead to multivalued energy
kernels in MR calculations \cite{(Dob07),(Dug09)}. Therefore, it is
interesting to go back to Skyrme's original idea and to consider a
zero-range gradientless three-body interaction. The form implemented
in the version {(v\codeversion)} of the code \pr{hfodd} is the one
defined in Refs.~\cite{(Sad13),(Sad13b)},

\begin{equation}
\label{eq:3body:gradientsless}
\hat{v}_3 (i,j,k) = 3 \, u_0 \,\delta(\mathbf r_{ij}) \, \delta(\mathbf r_{ik}) \,.
\end{equation}
In the literature, $t_3 = 3 \, u_0$ is often used for the coupling
constant instead \cite{(Vau72c),(Bei75b)}. Skyrme's
article~\cite{(Sky59b)} also proposed a gradientless four-body
contact interaction, which in some recent beyond mean-field
calculations~\cite{(Bal14d)} was used to complement the three-body
pseudopotential of Eq.~(\ref{eq:3body:gradientsless}). The form
implemented in the version {(v\codeversion)} of the code \pr{hfodd}
is again the one defined in Ref.~\cite{(Sad13)},

\begin{equation}
\label{eq:4body:gradientsless}
\hat{v}_4 (i,j,k,l)
= 12 \, v_0 \, \delta(\mathbf r_{ij}) \, \delta(\mathbf r_{ik})
 \, \delta(\mathbf r_{il}) \,.
\end{equation}
The factors three and twelve in Eqs.~(\ref{eq:3body:gradientsless})
and~(\ref{eq:4body:gradientsless}), respectively, count the number of different
permutations of the coordinates in the delta functions~\cite{(Sad13)}. The
contributions from these two terms in the particle-hole and like-particle
particle-particle channels of the energy density as detailed in
Ref.~\cite{(Sad13)} are fully implemented in version {(v\codeversion)} of the code \pr{hfodd}.

\subsection{Zero-range three-body gradient terms}
\label{subsec:gradient}

To allow for a greater flexibility of the three-body contribution to the EDF, one can also
consider contact interactions with gradients. Terms of this kind were occasionally
considered since the 1970s, see Ref.~\cite{(Sad13b)} for an overview, but were
up to now never used systematically. The most general isospin-invariant central
three-body pseudo-potential with two gradients can be written as \cite{(Sad13b)}
\begin{eqnarray}
\label{eq:3body:gradients}
\hat{v}(i,j,k)
= & \, u_1 \bigg\{
    \Big( 1  + y_1 \hat{P}^{\sigma}_{ij} \Big) \, \tfrac{1}{2} \,
    \Big[ \hat{\bm{k}}^{\,\dagger \, 2}_{ij} \, \delta(\bm{r}_{ik})  \, \delta(\bm{r}_{jk})
        + \delta(\bm{r}_{ik}) \, \delta(\bm{r}_{jk}) \, \hat{\bm{k}}^{\,2}_{ij} \Big]
     \nonumber \\
   & \; \; + \Big( 1  + y_1 \hat{P}^{\sigma}_{ik} \Big) \, \tfrac{1}{2} \,
    \Big[ \hat{\bm{k}}^{\,\dagger \, 2}_{ik} \, \delta(\bm{r}_{ij})  \, \delta(\bm{r}_{jk})
        + \delta(\bm{r}_{ij}) \, \delta(\bm{r}_{jk}) \, \hat{\bm{k}}^{\,2}_{ik} \Big]
      \nonumber \\
  & \; \; + \Big( 1  + y_1 \hat{P}^{\sigma}_{jk} \Big) \, \tfrac{1}{2} \,
    \Big[ \hat{\bm{k}}^{\,\dagger \, 2}_{jk} \, \delta(\bm{r}_{ij})  \, \delta(\bm{r}_{ik})
        + \delta(\bm{r}_{ij}) \, \delta(\bm{r}_{ik}) \, \hat{\bm{k}}^{\,2}_{jk} \Big]
    \bigg\}
              \nonumber \\
   & + u_2 \bigg\{\Big[ 1 + y_{21} \hat{P}^{\sigma}_{ij} + y_{22} \Big( \hat{P}^{\sigma}_{ik} + \hat{P}^{\sigma}_{jk} \Big) \Big]\,
      \hat{\bm{k}}^{\,\dagger}_{ij} \, \delta(\bm{r}_{ik}) \, \delta(\bm{r}_{jk}) \cdot \hat{\bm{k}}^{\,}_{ij}
              \nonumber \\
   &  \; \; + \Big[ 1 + y_{21} \hat{P}^{\sigma}_{ik} + y_{22} \Big( \hat{P}^{\sigma}_{ij} + \hat{P}^{\sigma}_{jk} \Big) \Big]\,
      \hat{\bm{k}}^{\,\dagger}_{ik} \, \delta(\bm{r}_{ij}) \, \delta(\bm{r}_{jk}) \cdot \hat{\bm{k}}^{\,}_{ik}
              \nonumber \\
   &  \; \; + \Big[ 1 + y_{21} \hat{P}^{\sigma}_{jk} + y_{22} \Big( \hat{P}^{\sigma}_{ij} + \hat{P}^{\sigma}_{ik} \Big) \Big]\,
      \hat{\bm{k}}^{\,\dagger}_{jk} \, \delta(\bm{r}_{ij}) \, \delta(\bm{r}_{ik}) \cdot \hat{\bm{k}}^{\,}_{jk}
     \bigg\}
      \, ,
\end{eqnarray}
with the five parameters $u_1$, $y_1$, $u_2$, $y_{21}$ and $y_{22}$.
The EDF resulting from the three-body contact generators can be expressed as an integral over a local
energy density that is built out of the same normal and pairing densities as the standard Skyrme EDF.
Contributions from the three-body contact generators to the particle-hole and like-particle $T=1$
particle-particle terms in the EDF, as elaborated in Ref.~\cite{(Sad13b)}, are fully implemented in the
version {(v\codeversion)} of the code \pr{hfodd}.

\subsection{Zero-range tensor terms}
\label{subsec:tensor}

The Skyrme interaction, in a majority of practical implementations,
does not include tensor terms. In version {(v\codeversion)} of the code \pr{hfodd}, we
implemented  the conventional zero-range tensor interaction
considered already by Skyrme in his seminal work~\cite{(Sky59b)}, see
also
Refs.~\cite{(Sta77a),(Per04c),(Les07b),(Zal08a),(Ben09d),(Hel12)} and
references therein:
\begin{eqnarray}
\label{eq:tensor}
\hat{v}^{\rm{T}}(i,j)  & = &  \tfrac{1}{2} t_e
\Big\{ \left(
3 (\bm{\sigma}_i \cdot  \bm{k}' ) (\bm{\sigma}_j \cdot  \bm{k}' )  -  (\bm{\sigma}_i \cdot   \bm{\sigma}_j )   \bm{k}'^2
\right)  \delta\left( \gras{r}_{ij} \right)   \notag \\
&& \qquad + \delta\left( \gras{r}_{ij} \right) \left(
3 (\bm{\sigma}_i \cdot  \bm{k} ) (\bm{\sigma}_j \cdot  \bm{k} )  -  (\bm{\sigma}_i \cdot   \bm{\sigma}_j )  \bm{k} ^2
\right)   \Big\} \notag \\
& + & t_o \Big\{ \tfrac{3}{2} (\bm{\sigma}_i \cdot  \bm{k}' ) \delta\left( \gras{r}_{ij} \right)  (\bm{\sigma}_j \cdot  \bm{k} )
               +\tfrac{3}{2} (\bm{\sigma}_j \cdot  \bm{k}' ) \delta\left( \gras{r}_{ij} \right)  (\bm{\sigma}_i \cdot  \bm{k} )   \notag \\
&& \qquad - (\bm{\sigma}_i \cdot   \bm{\sigma}_j ) \bm{k}'   \delta\left( \gras{r}_{ij} \right)  \bm{k} \Big\},
\end{eqnarray}
where the first term acts in the relative S- and D-waves whereas the
second component acts in the P-wave. Parameters $t_e$ and $t_o$
are new low-energy coupling constants (LECs) which have to be
adjusted to experimental data.

\newcommand{\myrm}{}
The contact tensor interaction contributes to both the time-even
(bilinear in time-even densities) and time-odd (bilinear in time-odd
densities) terms of the local EDF. The tensor part of the generalized
Skyrme EDF in the time-even (t-even) sector is:
\begin{eqnarray}\label{eq:T_TE}
{\cal H}_t^{{\rm T,\, t-even}}  \left( \gras{r} \right)  &=& C_t^{\myrm J}  {\sf J}_t^2   +
\Delta {\cal H}_t^{{\rm T,\, t-even}}  \left( \gras{r} \right)  \nonumber \\
 &=& (C_t^{\myrm J} + B_t^{\myrm J} ) {\sf J}_t^2
  + B_t^{\myrm X} \tfrac{1}{2} \Big\{ \Big(\Tr {J}_t \Big)^2 + \Tr {J}_t^2 \Big\},
\end{eqnarray}
where $t=0,1$ denotes isoscalar and isovector densities,
respectively.  Within the conventional $pn$-separable DFT these are
simply sums and differences of the neutron and proton densities,
respectively. The standard local
spin-current pseudotensor density ${J}_{t, \mu\nu}( \gras{r})$
is defined through the nonlocal spin density ${s}_{t, \nu} (\gras{r}, \gras{r}')$ as:
\begin{equation}\label{eq:J}
{J}_{t, \mu\nu} ( \gras{r}) = \tfrac{1}{2i} \big\{ ({\nabla}_\mu - {\nabla}'_\mu )
{s}_{t, \nu} (\gras{r}, \gras{r}') \big\}_{\gras{r} =\gras{r}'}
\end{equation}
with the sum of squares of its components conventionally denoted as ${\sf J}_t^2$,
\begin{equation}\label{eq:J2}
{\sf J}_t^2 \equiv \sum_{\mu \nu} J_{t, \mu\nu}^2 .
\end{equation}
The tensor part of the generalized
Skyrme EDF in the time-odd (t-odd) sector is:
\begin{eqnarray}\label{eq:tensor_TO}
{\cal H}_t^{{\rm T,\, t-odd}}  \left( \gras{r} \right)
&=& C_t^{\myrm T} \bm{s}_t \cdot \bm{T}_t  + C_t^{\Delta s} \bm{s}_t \cdot \Delta \bm{s}_t  +
\Delta {\cal H}_t^{{\rm T,\, t-odd}} \left( \gras{r} \right) \nonumber \\
&=& (C_t^{\myrm T} + B_t^{\myrm T}) \bm{s}_t \cdot \bm{T}_t
 +  (C_t^{\Delta s} + B_t^{\Delta s})\bm{s}_t \cdot \Delta \bm{s}_t  \nonumber \\
&+& B_t^{\myrm F} \bm{s}_t \cdot \bm{F}_t  + B_t^{\nabla s} (\bm{\nabla} \cdot \bm{s}_t )^2,
\end{eqnarray}
where ${\bm s}$ and ${\bm T}$ are the standard spin and spin-kinetic densities,
respectively, and the tensor-kinetic
density~\cite{(Per04c)} ${\bm F}$ reads,
\begin{equation}
F_{t,\nu} ( \gras{r}) = \tfrac{1}{2} \Big[\sum_\mu \Big\{
({\nabla}_\nu {\nabla}'_\mu  + {\nabla}'_\nu {\nabla}_\mu )
s_{t,\mu} (\gras{r}, \gras{r}') \Big\} \Big]_{\gras{r} =\gras{r}'}.
\end{equation}

In Eqs.~(\ref{eq:T_TE}) and (\ref{eq:tensor_TO}),  $C_t^{\myrm J}$,
$C_t^{\myrm T}$, and $C_t^{\Delta s}$ denote the functional tensor coupling
constant due to the central field, as defined in
Refs.~\cite{(Dob95b),(Dob97b)}. Tensor interaction (\ref{eq:tensor})
adds to these terms its own contributions, $B_t^{\myrm J}$, $B_t^{\myrm
T}$, and $B_t^{\Delta s}$, respectively, but these additions do not change
the structure of the functional. However, the tensor force also adds
new terms of the functional specified by terms $\Delta {\cal H}_t^{{\rm T,\, t-even}}$
and $\Delta {\cal H}_t^{{\rm T,\, t-odd}}$, and by coupling constants
$B_t^{\myrm X}$,$B_t ^{\myrm F}$, and $B_t^{\nabla s}$.
The new functional coupling constants
$B_t^{\myrm X}$, $B_t^{\myrm F}$, and  $B_t^{\nabla s}$ for $t=0,1$, are encoded under
the names \tv{CSCT\_X}, \tv{CKIT\_X}, and  \tv{CSPT\_X} for \tv{X=P,M}, respectively, and
printed in the code's output.

The twelve new coupling constants $B_t^{\myrm J}$, $B_t^{\myrm X}$, $B_t^{\myrm
T}$, $B_t^{\Delta s}$, $B_t^{\myrm F}$, and  $B_t^{\nabla s}$ for $t=0,1$
relate to the two parameters of the tensor interaction $t_e$ and $3t_o$ as
\begin{eqnarray}\label{eq:tensor_CC}
B_0^{\myrm J} &=&  \tfrac{1}{8} (t_e + 3t_o) , \quad~\, B_1^{\myrm J}  = - \tfrac{1}{8} (t_e - t_o), \\
B_0^{\Delta s} &=&  \tfrac{3}{32} (t_e - t_o ), \quad B_1^{\Delta s} =  - \tfrac{1}{32} (3t_e + t_o ),
 \quad  B_t^{\nabla s} = 3 B_t^{\Delta s} \\
B_t^{\myrm X}  &=& - 3B_t^{\myrm J}, \quad \quad~~~ B_t^{\myrm T} = - B_t^{\myrm J} ,
 \quad \quad \quad B_t^{\myrm F} = 3 B_t^{\myrm J}.
\end{eqnarray}

In order to elucidate the role of the last term on the
right-hand-side of Eq.~(\ref{eq:T_TE}) it is convenient to decompose
the spin-current pseudotensor density (\ref{eq:J}) into pseudoscalar, vector, and
rank-2 pseudotensor components~\cite{(Per04c)}, which gives Eq.~(\ref{eq:J2}) in the form:
\begin{equation}
{\sf J}_t^2 = \tfrac{1}{3} \big( J_t^{\rm (0)} \big)^2 + \tfrac{1}{2} {\bm J_t}^2 + \sum_{\mu\nu} \Big( J_{t,\mu\nu}^{\rm (2)} \Big)^2.
\end{equation}
Moreover, since:
\begin{equation}
\tfrac{1}{2} \Big\{ \Big(\sum_\mu J_{t, \mu\mu} \Big)^2
+ \sum_{\mu \nu} J_{t, \mu\nu} J_{t, \nu\mu} \Big\} = \tfrac{2}{3} \big( J_t^{\rm (0)} \big)^2 - \tfrac{1}{4} {\bm J_t}^2 + \tfrac{1}{2}
\sum_{\mu\nu} \Big( J_{t,\mu\nu}^{\rm (2)} \Big)^2,
\end{equation}
the contribution $\Delta {\cal H}_t^{{\rm T,\, t-even}}  \left(
\gras{r} \right)$ can  be rewritten to the following form (see
Ref.~\cite{(Per04c)}) :
\begin{equation}
\Delta {\cal H}_t^{{\rm T, \, t-even}} =
B_t^{J_0} \big( J_t^{\rm (0)} \big)^2 + B_t^{J_1} {\bm J_t}^2 + B_t^{J_2} \Big( J_{t,\mu\nu}^{\rm (2)} \Big)^2,
\end{equation}
where
\begin{equation}
B_t^{J_0} = -\tfrac{5}{3} B_t^{\myrm{J}}, \quad
B_t^{J_1} =  \tfrac{5}{4} B_t^{\myrm{J}}, \quad {\rm and} \quad
B_t^{J_2} = -\tfrac{1}{2} B_t^{\myrm{J}}.
\end{equation}
It means that the contact zero-range tensor force does not introduce
new terms in the time-even part of the Skyrme functional. It only
modifies the conventional Skyrme EDF coupling constants $C_t^{J_0} =
\tfrac{1}{3} C_t^{J}$,  $C_t^{J_1} = \tfrac{1}{2} C_t^{J}$, and $C_t^{J_2}=
C_t^{J}$~\cite{(Per04c)}. Hence, within the single-reference DFT  the effect of the
tensor interaction can be mimicked by readjusting the Skyrme-force
values of $C_t^{J_0}$,  $C_t^{J_1}$, and $C_t^{J_2}$ what explains
why the tensor interaction is often neglected within the standard
Skyrme force. In particular, in version {(v\codeversion)} of the code \pr{hfodd},
separate use of the coupling constants $C_t^{J_0}$,  $C_t^{J_1}$, and $C_t^{J_2}$
has not yet been implemented. Nevertheless, a readjustment or, in  general,  a
direct fit of  functional's coupling constants to a dedicated set of
empirical data is well within the spirit of the single-reference DFT, which treats EDF
as a primary physical object~\cite{(Les07b),(Ben09d),(Hel12)}. One should bear in mind, however, that
it breaks bonds between the  functional and the underlying
interaction what precludes its application within the multi-reference
extensions due to singularities that appear in the energy
kernels~\cite{(Dob07),(Lac09),(Ben09),(Sat14g)}.

Values of the tensor LECs, $t_e$ and $t_o$, can be determined through
the large-scale multi-parameter fit to masses, see
e.g.~\cite{(Les07b),(Ben09d)}. An alternative way was proposed in
Ref.~\cite{(Zal08a)}. This method is based on simultaneous fit of the
spin-orbit strength and tensor's LECs to the single-particle levels
and spin-orbit splittings in double-magic nuclei $^{40}$Ca,
$^{48}$Ca, and $^{56}$Ni.

\subsection{Zero-range isospin-breaking terms}
\label{subsec:ISB}

Isospin symmetry is not a fundamental symmetry of nature. At the fundamental level of quantum
chromodynamics, it is broken due to different masses and charges of constituent quarks. At the energy scales typical
for nuclear physics, where quarks and gluons  are not resolvable and the proper degrees of freedom
are point-like nucleons, the isospin-symmetry breaking (ISB) comes, predominantly, from the long-range Coulomb
interaction and, albeit to a much lesser degree, from the short-range effective ISB nuclear forces.

The  effective nuclear force can be divided into four different  classes following the scheme introduced
by Henley and Miller~\cite{(Mil95a),(Hen79a)}.  Apart of the dominant class-I isoscalar force, there are three different classes
of the ISB forces including class-II isotensor force, class-III isovector force, and
class~IV force which mixes isospin already at the two-body level.
In finite nuclei the ISB effects manifest themselves very clearly already in the simplest
observables, the nuclear masses, through  the mirror displacement energy (MDE):
\begin{equation}
\mathrm{MDE}=BE(T,T_z\mbox{=}-T)-BE(T,T_z\mbox{=}+T), \label{eq:MDE}
\end{equation}
and triplet displacement energies (TDE):
\begin{equation}
\mathrm{TDE}=BE(T\mbox{=}1,T_z\mbox{=}-1)+BE(T\mbox{=}1,T_z\mbox{=}+1)
-2BE(T\mbox{=}1,T_z\mbox{=}0). \label{eq:TDE}
\end{equation}
The MDEs and TDEs are almost exclusively sensitive to the charge symmetry breaking (CSB  or class-III)
and charge independence breaking (CIB  or class-II) terms in the nuclear Hamiltonian, respectively.
Class-IV force will be neglected as no firm evidence of the effects related to this force was identified so far in
many-body data.

It is well known, that none of the displacement energies  (\ref{eq:MDE}) or (\ref{eq:TDE}) can be reproduced using models involving Coulomb
interaction as the only source of the ISB, see Refs.~\cite{(Nol69),(Orm89),(Col98),(Bro00c),(Kan13),(Orm17),(Bac18),(Roc18a),(Bac19)}
and references cited therein. This deficiency concerns, in particular, the nuclear DFT including
its most popular realization based on  Skyrme forces, which are isoscalar by construction.

In order to account for the MDEs and TDEs we extended the conventional Skyrme
interaction by adding, first, the leading-order (LO) contact interactions of class-II and class-III
\cite{(Bac18)} and, subsequently, generalizing the ISB Skyrme interaction to the next-to-leading (NLO)
order in gradient  expansion~\cite{(Bac19)}. The introduced ISB terms read:
 \begin{eqnarray}
\hat{V}^{\rm{II}}(i,j)  & =  &
t_0^{\rm{II}}   \left( 1 + x_0^\mathrm{II} \hat{P}_{ij}^\sigma \right) \delta\left(\gras{r}_{ij} \right) \hat T^{(ij)} \nonumber \\
 & + & \bigg[ \, \tfrac{1}{2} t_1^{\rm{II}}  \left( 1 + x_1^\mathrm{II} \hat{P}_{ij}^\sigma \right)
\left( \delta\left( \gras{r}_{ij} \right) \bm{k}^2 + \bm{k}'^2 \delta\left(\gras{r}_{ij} \right) \right) \nonumber \\
 &&+
t_2^{\rm{II}}  \left( 1 + x_2^\mathrm{II} \hat{P}_{ij}^\sigma \right)
\bm{k}' \delta\left(\gras{r}_{ij} \right) \bm{k} \, \bigg]  \hat T^{(ij)} ,
\label{eq:classII_NLO} \\
\hat{V}^{\rm{III}}(i,j)  &  =  &
t_0^{\rm{III}}   \left( 1 + x_0^\mathrm{III} \hat{P}_{ij}^\sigma \right) \delta\left(\gras{r}_{ij} \right) \hat T_z^{(ij)} \nonumber \\
 & + & \bigg[ \, \tfrac{1}{2} t_1^{\rm{III}}  \left( 1 + x_1^\mathrm{III} \hat{P}_{ij}^\sigma \right)
\left( \delta\left( \gras{r}_{ij} \right) \bm{k}^2 + \bm{k}'^2 \delta\left(\gras{r}_{ij} \right) \right) \nonumber \\
&&+  t_2^{\rm{III}}  \left( 1 + x_2^\mathrm{III} \hat{P}_{ij}^\sigma \right)
\bm{k}' \delta\left(\gras{r}_{ij} \right) \bm{k} \, \bigg]  \hat T_z^{(ij)}, \nonumber \\
\label{eq:classIII_NLO}
\end{eqnarray}
where $\hat{P}_{ij}^\sigma$ stands for the spin-exchange operator,
$\gras{r}_{ij} = \gras{r}_i - \gras{r}_j$,
$\bm{k}  =  \tfrac{1}{2i}\left(\bm{\nabla}_i-\bm{\nabla}_j\right)$ and
$\bm{k}' = -\tfrac{1}{2i}\left(\bm{\nabla}_i-\bm{\nabla}_j\right)$ are the standard relative-momentum
operators acting to the right and left, respectively, whereas
$\hat T^{(ij)} = 3\hat{\tau}_3^{(i)}\hat{\tau}_3^{(j)}-\hat{\vec{\tau}}^{(i)}\circ\hat{\vec{\tau}}^{(j)}$ and
$\hat T_z^{(ij)} = \hat{\tau}_3^{(i)}+\hat{\tau}_3^{(j)}$ are the isotensor and isovector operators.
The contributions to energy density functional (EDF)  from the isovector and isotensor forces read:
\begin{eqnarray}
\mathcal{H}_{{\rm NLO}}^{\rm III} & = &
 \tfrac{1}{2} t_0^{\rm III} \left( 1 - x_0^{\rm III} \right)
\Big(
        \rho_{n}^2
        - \rho_{p}^2
        - \bm{s}_{n}^2
        + \bm{s}_{p}^2
\Big) \nonumber \notag \\
 & + & \tfrac{1}{4} t_1^{\rm III} \left( 1 - x_1^{\rm III} \right)
\Big(
                \tau_{n}
                \rho_{n}
                -
                \tau_{p}
                \rho_{p}
                -
                \bm{T}_{n} \cdot
                \bm{s}_{n}
                +
                \bm{T}_{p} \cdot
                \bm{s}_{p}
\Big) \nonumber \notag\\
& + &
\tfrac{1}{4} t_2^{\rm III} \left( 1 + x_2^{\rm III} \right)
\Big(
                3
                \tau_{n}
                \rho_{n}
                - 3
                \tau_{p}
                \rho_{p}
                +
                \bm{T}_{n} \cdot
                \bm{s}_{n}
                -
                \bm{T}_{p} \cdot
                \bm{s}_{p}
\Big)
\nonumber  \notag \\
 & - & \tfrac{3}{16} t_1^{\rm III} \left( 1 - x_1^{\rm III} \right)
\Big(
                \Delta\rho_{n}
                \rho_{n}
                -
                \Delta\rho_{p}
                \rho_{p}
                -
                \Delta\bm{s}_{n} \cdot
                \bm{s}_{n}
                +
                \Delta\bm{s}_{p} \cdot
                \bm{s}_{p}
\Big) \nonumber \notag\\
& + & \tfrac{1}{16} t_2^{\rm III} \left( 1 + x_2^{\rm III} \right)
\Big(
                3
                \Delta\rho_{n}
                \rho_{n}
                - 3
                \Delta\rho_{p}
                \rho_{p}
                +
                \Delta\bm{s}_{n} \cdot
                \bm{s}_{n}
                -
                \Delta\bm{s}_{p} \cdot
                \bm{s}_{p}
\Big) \nonumber \notag\\
& - & \tfrac{1}{4} t_1^{\rm III} \left( 1 - x_1^{\rm III} \right)
\Big(
        \bm{j}_{n}^2
        - \bm{j}_{p}^2
        - \bm{J}_{n}^2
        + \bm{J}_{p}^2
\Big) \nonumber \notag\\
& - & \tfrac{1}{4} t_2^{\rm III} \left( 1 + x_2^{\rm III} \right)
\Big(
        3\bm{j}_{n}^2
        - 3\bm{j}_{p}^2
        + \bm{J}_{n}^2
        - \bm{J}_{p}^2
\Big), \label{eq:edf_III}
\end{eqnarray}
\begin{eqnarray}
\mathcal{H}_{{\rm NLO}}^{\rm II} & = &
\tfrac{1}{2} t_0^{\rm II} \left( 1 - x_0^{\rm II} \right)
\Big(
                \rho_{n}^2
                +
                \rho_{p}^2
                - 2
                \rho_{n}
                \rho_{p}
                - 2
                \rho_{np}
                \rho_{pn}
 \nonumber \\ &&
                -
                \bm{s}_{n}^2
                -
                \bm{s}_{p}^2
                + 2
                \bm{s}_{n} \cdot
                \bm{s}_{p}
                + 2
                \bm{s}_{np} \cdot
                \bm{s}_{pn}
\Big) \nonumber \\
 & +& \tfrac{1}{4} t_1^{\rm II} \left( 1 - x_1^{\rm II} \right)
\Big(
                \tau_{n}
                \rho_{n}
                +
                \tau_{p}
                \rho_{p}
                -
                \tau_{n}
                \rho_{p}
                -
                \tau_{p}
                \rho_{n}
                -
                \tau_{np}
                \rho_{pn}
                -
                \tau_{pn}
                \rho_{np} \notag\\
& &
                -
                \bm{T}_{n} \cdot
                \bm{s}_{n}
                -
                \bm{T}_{p} \cdot
                \bm{s}_{p}
                +
                \bm{T}_{n} \cdot
                \bm{s}_{p}
                +
                \bm{T}_{p} \cdot
                \bm{s}_{n}
                +
                \bm{T}_{np} \cdot
                \bm{s}_{pn}
                +
                \bm{T}_{pn} \cdot
                \bm{s}_{np}
\Big)
\nonumber \notag\\
& + & \tfrac{1}{4} t_2^{\rm II} \left( 1 + x_2^{\rm II} \right)\!
\Big(
                3\tau_{n}
                \rho_{n}
                +
                3\tau_{p}
                \rho_{p}
                -
                3\tau_{n}
                \rho_{p}
                -
                3\tau_{p}
                \rho_{n}
                -3
                \tau_{np}
                \rho_{pn}
                -3
                \tau_{pn}
                \rho_{np} \nonumber\notag\\
& &
                +
                \bm{T}_{n} \cdot
                \bm{s}_{n}
                +
                \bm{T}_{p} \cdot
                \bm{s}_{p}
                -
                \bm{T}_{n}
                \cdot \bm{s}_{p}
                -
                \bm{T}_{p}
                \cdot \bm{s}_{n}
                -
                \bm{T}_{np} \cdot
                \bm{s}_{pn}
                -
                \bm{T}_{pn} \cdot
                \bm{s}_{np}
\Big) \nonumber \notag \\
 &  - & \tfrac{3}{16} t_1^{\rm II} \left( 1 - x_1^{\rm II} \right)
\Big(
                \Delta\rho_{n}
                \rho_{n}
                +
                \Delta\rho_{p}
                \rho_{p}
                -
                \Delta\rho_{n}
                \rho_{p}
 \nonumber \\ &&
                -
                \Delta\rho_{p}
                \rho_{n}
                -
                \Delta\rho_{np}
                \rho_{pn}
                -
                \Delta\rho_{pn}
                \rho_{np} \nonumber \notag\\
& &
                -
                \Delta\bm{s}_{n} \cdot
                \bm{s}_{n}
                -
                \Delta\bm{s}_{p} \cdot
                \bm{s}_{p}
                +
                \Delta\bm{s}_{n} \cdot
                \bm{s}_{p}
 \nonumber \\ &&
                +
                \Delta\bm{s}_{p} \cdot
                \bm{s}_{n}
                +
                \Delta\bm{s}_{np} \cdot
                \bm{s}_{pn}
                +
                \Delta\bm{s}_{pn} \cdot
                \bm{s}_{np}
\Big) \nonumber \notag\\
& + &\tfrac{1}{16} t_2^{\rm II} \left( 1 + x_2^{\rm II} \right)
\Big(
                3\Delta\rho_{n}
                \rho_{n}
                +
                3\Delta\rho_{p}
                \rho_{p}
                -
                3\Delta\rho_{n}
                \rho_{p}
 \nonumber \\ &&
                -
                3\Delta\rho_{p}
                \rho_{n}
                -3
                \Delta\rho_{np}
                \rho_{pn}
                -3
                \Delta\rho_{pn}
                \rho_{np} \nonumber \notag\\
& &
                +
                \Delta\bm{s}_{n} \cdot
                \bm{s}_{n}
                +
                \Delta\bm{s}_{p} \cdot
                \bm{s}_{p}
                -
                \Delta\bm{s}_{n}
                \cdot \bm{s}_{p}
 \nonumber \\ &&
                -
                \Delta\bm{s}_{p}
                \cdot \bm{s}_{n}
                -
                \Delta\bm{s}_{np} \cdot
                \bm{s}_{pn}
                -
                \Delta\bm{s}_{pn} \cdot
                \bm{s}_{np}
\Big)
\nonumber \notag\\
& - &\tfrac{1}{4} t_1^{\rm II} \left( 1 - x_1^{\rm II} \right)
\Big(
                \bm{j}_{n}^2
                +
                \bm{j}_{p}^2
                - 2
                \bm{j}_{n} \cdot
                \bm{j}_{p}
                - 2
                \bm{j}_{np} \cdot
                \bm{j}_{pn}
 \nonumber \\ &&
                -
                \bm{J}_{n}^2
                -
                \bm{J}_{p}^2
                + 2
                \bm{J}_{n} \cdot
                \bm{J}_{p}
                + 2
                \bm{J}_{np} \cdot
                \bm{J}_{pn}
\Big)
\nonumber \notag\\
& - & \tfrac{1}{4} t_2^{\rm II} \left( 1 + x_2^{\rm II} \right)
\Big(
                3\bm{j}_{n}^2
          +
                3\bm{j}_{p}^2
                -
                6\bm{j}_{n} \cdot
                \bm{j}_{p}
                -6
                \bm{j}_{np} \cdot
                \bm{j}_{pn}
 \nonumber \\ &&
                +
                \bm{J}_{n}^2
                +
                \bm{J}_{p}^2
                -2
                \bm{J}_{n}
                \cdot \bm{J}_{p}
                -2
                \bm{J}_{np} \cdot
                \bm{J}_{pn}
\Big),  \label{eq:edf_II}
\end{eqnarray}
where  $\rho, \tau, {\bm s}, {\bm T}, {\bm j},$ and
${\bm J}$ denote  the standard particle, kinetic, spin, spin-kinetic, current, and vector spin-current
densities, respectively. It is of importance to underline, that the contributions
of class-III local force  (\ref{eq:edf_III}) depend on the standard $nn$ and $pp$ densities.
This force can be, therefore, taken into account within the conventional $pn$-separable DFT
approach. In contrast, contributions due to the class-II force (\ref{eq:edf_II}) depend explicitly on the mixed
$pn$-densities and require the use of $pn$-mixed DFT formulated in Refs.~\cite{(Sat13d),(She14a)} and
implemented in the previous version (v2.73y) of code \pr{hfodd}~\cite{(Sch17d)} together with the isocranking method~\cite{(Sat01a),(Sat01b),(Sat13d)},
which allows control of the isospin degree of freedom.

Inclusion of spin-exchange terms in the ISB contact forces (\ref{eq:classII_NLO}) and
(\ref{eq:classIII_NLO}) leads to a simple rescaling of the $t_i^{\rm II}$ and $t_i^{\rm III}$ (for $i$=0, 1, and 2)
parameters. Hence, the spin-exchange terms are effectively redundant and can be omitted
from the ISB contact forces. Hence, the LO (NLO) ISB forces introduce two (six) new LECs, which must be adjusted to the existing data. Our studies~\cite{(Bac18),(Bac19)}
show that the fitting strategy for the new LECs can be considerably simplified for the following reasons:
Firstly,  for the physically relevant values of the ISB LECs, contributions of the ISB terms to the total binding energies appear to be
relatively small (at least this can be concluded from studies performed in the vicinity of the $N=Z$ line). This implies that the ISB terms can be treated
as small perturbations to the Skyrme force with frozen parameters. Secondly, the MDEs and TDEs
are almost exclusively sensitive to the CSB and CIB  terms in the nuclear Hamiltonian, respectively.
As a consequence,  the isovector and isotensor
LECs can be adjusted separately through the global fit to the experimental MDEs and TDEs, respectively.
Such a strategy was applied to the SLy4~\cite{(Cha98a)}, and SkM$^*$~\cite{(Bar82b)} forces as well as to the
two variants, SV$_{\rm T}$ and SV$_{\rm T,\, SO}$ (see Refs.~\cite{(Sat14g),(Kon16)}), of the SV density-independent Skyrme
interaction~\cite{(Bei75b)} that can be used in the beyond-mean-field multi-reference
DFT calculations.  The resulting parameters are collected in Table~\ref{tab:LECS}. They seem to be consistent
with the values given in Ref.~\cite{(Roc18a)}, where a different fitting strategy was used. The adjusted LECs lead to
an excellent global description of the existing data on MDEs and TDEs in the isospin doublets and triplets,
see Refs.~\cite{(Bac18),(Bac19)} for further details.
\begin{table}[t!]
\centering
\renewcommand{\arraystretch}{1.2}
\begin{tabular}{c||c|l|l||c|l|l}
\multirow{2}{*}{Skyrme force}   &       \multicolumn{3}{c||}{LO approximation}                  &       \multicolumn{3}{c}{NLO approximation}                   \\\cline{2-7}
        &       $t$     &       \multicolumn{1}{c|}{class-II}   &       \multicolumn{1}{c||}{class-III} &       $t$     &       \multicolumn{1}{c|}{class-II}   &       \multicolumn{1}{c}{class-III}   \\\hline\hline
SV$_{\rm T}$    &       $t_0$   &       $3.7\pm0.4$     &       $-7.3\pm0.3$    &       $t_0$   &       $-16\pm3$       &       $\phantom{-}11\pm2$     \\
        &               &               &               &       $t_1$   &       $\phantom{-}22\pm3$     &       $-14\pm4$       \\
        &               &               &               &       $t_2$   &       $\phantom{-0}1\pm1$     &       \phantom{0}$-7.8\pm0.8$ \\\hline
SV$_{\rm T,\, SO}$      &       $t_0$   &               &       $-6.7\pm0.3$    &       $t_0$   &               &       $\phantom{-}5\pm2$      \\
        &               &               &               &       $t_1$   &               &       $-3\pm3$        \\
        &               &               &               &       $t_2$   &               &       $-7.4\pm0.7$    \\\hline
SkM$^*$ &       $t_0$   &       $5.2\pm0.8$     &       $-5.4\pm0.2$    &       &               &               \\\hline
SLy4    &       $t_0$   &       $5.1\pm0.8$     &       $-5.5\pm0.2$    &       &               &               \\
\end{tabular}
\caption{
Values of LECs for LO and NLO ISB forces adjusted to the experimental
data for the following parameterizations of the (isoscalar) Skyrme
forces: SV$_\mathrm{T}^\mathrm{ISB}$,
SV$_\mathrm{T,\,SO}^\mathrm{ISB}$, SkM*$^\mathrm{ISB}$ i
SLy4$^\mathrm{ISB}$. The $t_0$ parameter is given in MeV\,fm$^3$,
whereas the  $t_1$ and $t_2$ parameters are in  MeV\,fm$^5$.}
\label{tab:LECS}
\end{table}

\subsection{Finite-range higher-order regularized terms}
\label{subsec:Regularized}

Following the notation introduced in Ref.~\cite{(Rai14a)}, see also Refs.~\cite{(Dob12b),(Ben14a),(Dav13a)},
we define the Cartesian form
of the (non-anti\-symmetrized) central
pseudopotential as
\begin{eqnarray}
\label{eq:1}
 \mathcal{V}_C(\bm{r}'_1\bm{r}'_2;\bm{r}_1\bm{r}_2) &=& \sum_{nj}
   \left(W^{(n)}_j \hat{1}_\sigma\hat{1}_\tau+B^{(n)}_j \hat{1}_\tau  \hat{P}^\sigma
        -H^{(n)}_j \hat{1}_\sigma\hat{P}^\tau-M^{(n)}_j \hat{P}^\sigma\hat{P}^\tau\right)
\nonumber \\
&&~~~~~~\times \hat{O}^{(n)}_j(\bm{k}',\bm{k})\delta(\bm{r}'_1-\bm{r}_1)\delta(\bm{r}'_2-\bm{r}_2)
g_a(\bm{r}_1-\bm{r}_2),
\end{eqnarray}
which contains the standard identity ($\hat{1}_{\sigma,\tau}$) and
exchange ($\hat{P}^{\sigma,\tau}$) operators in the spin and isospin
spaces and a Gaussian formfactor,
\begin{eqnarray}
\label{eq:1a}
g_a(\bm{r})&=&\frac{e^{-{\bm{r}^2}/{a^2}}}{\left(a\sqrt{\pi}\right)^3},
\end{eqnarray}
defined by its width $a$.
In Eq.~(\ref{eq:1}), index $n=0,2,\ldots$ denotes the order of
differential operator $\hat{O}^{(n)}_j(\bm{k}',\bm{k})$, index $j=1,2,\ldots$
numbers different operators of the same order, and the relative-momentum
operators are defined as  $\bm{k}=(\bm{\nabla}_1-\bm{\nabla}_2)/2i$
and  $\bm{k}'=(\bm{\nabla}'_1-\bm{\nabla}'_2)/2i$.
The standard Wigner, Bartlett, Heisenberg, and Majorana coupling constants,
$W^{(n)}_j$,
$B^{(n)}_j$,
$H^{(n)}_j$, and
$M^{(n)}_j$,
can also be expressed by the strength parameters
$t^{(n)}_j$,
$x^{(n)}_j$,
$y^{(n)}_j$, and
$z^{(n)}_j$ as
\begin{eqnarray}
\label{eq:2}
 &&
W^{(n)}_j=t^{(n)}_j          ,\quad
B^{(n)}_j=t^{(n)}_j x^{(n)}_j,\quad
H^{(n)}_j=t^{(n)}_j y^{(n)}_j,\quad
M^{(n)}_j=t^{(n)}_j z^{(n)}_j.
\end{eqnarray}

Up to sixth order ($n=6$), a full classification of operators
$\hat{O}^{(n)}_j(\bm{k}',\bm{k})$ was presented in Eqs.~(39)--(54) of
Ref.~\cite{(Rai14a)}. The EDF generated from the pseudo-potential
functional generators for both the particle and pairing channels can
be found in \cite{(Ben17a)}. All fields are implemented
self-consistently in the present version {(v\codeversion)} of the
code \pr{hfodd} and the computation of
the HO-basis spatial matrix elements of pseudopotentials is coded
using the following integrated-by-parts form,
\begin{eqnarray}
\label{eq:3}
\langle\bm{n}'_1\bm{n}'_2|\hat{O}^{(n)}_j|\bm{n}_1\bm{n}_2\rangle
 &\equiv&
\int\rmd\bm{r}'_1\rmd\bm{r}'_2\rmd\bm{r}_1\rmd\bm{r}_2\,
\delta(\bm{r}'_1-\bm{r}_1)\delta(\bm{r}'_2-\bm{r}_2)g_a(\bm{r}_1-\bm{r}_2)
\nonumber \\ &&\times\,
\hat{O}^{(n)}_j(\bm{k}',\bm{k})
\psi_{\bm{n}'_1}(\bm{r}'_1)\psi_{\bm{n}'_2}(\bm{r}'_2)
\psi_{\bm{n} _1}(\bm{r} _1)\psi_{\bm{n} _2}(\bm{r} _2),
\end{eqnarray}
where $\psi_{\bm{n}}(\bm{r})\equiv\psi_{n_xn_yn_z}(x,y,z)$ are the 3D
deformed HO wave functions, Eqs.~(I-71)--(I-73)~\cite{(Dob97b)}.

Using the explicit expressions~\cite{(Rai14a)} for differential operators $\hat{O}^{(n)}_j(\bm{k}',\bm{k})$,
one can tediously but straightforwardly rewrite them as sums of terms that are products of
differential operators acting in the $x$, $y$, and $z$ directions,
\begin{eqnarray}
\label{eq:4}
\hat{O}^{(n)}_j(\bm{k}',\bm{k})&=&\sum_{\bm{m},\bm{l}}K^{(n;\bm{m})}_{j;\bm{l}}
\hat{O}^{(m_x)}_{l_x}(k_x',k_x)
\hat{O}^{(m_y)}_{l_y}(k_y',k_y)
\hat{O}^{(m_z)}_{l_z}(k_z',k_z),
\end{eqnarray}
where $K^{(n;\bm{m})}_{j;\bm{l}}\equiv K^{(n;m_x,m_y,m_z)}_{j;l_x,l_y,l_z}$
for $m_x+m_y+m_z=n$ are integer coefficients and 1D
differential operators $\hat{O}^{(m)}_l(k',k)$ of order $m=0,2,\ldots$ and index
$l=0,1,\ldots,m/2$ are given by
\begin{eqnarray}
\label{eq:5}
\hat{O}^{(m)}_l(k',k)&\equiv&
\left((k'\,^2+k^2)/2\right)^{l}
\left(k'k\right)^{m/2-l}
 \nonumber \\ &&
=2^{-l}\sum_{l'=0}^l \newton{l}{l'}k'\,^{m/2+2l'-l}k^{m/2-2l'+l}.
\end{eqnarray}
The matrix element given in Eq.~(\ref{eq:3}) is thus equal to the sum of products
of 1D matrix elements $\langle{n}'_1{n}'_2|\hat{O}^{(m)}_l|{n}_1{n}_2\rangle$,
\begin{eqnarray}
\label{eq:6}
\langle\bm{n}'_1\bm{n}'_2|\hat{O}^{(n)}_j|\bm{n}_1\bm{n}_2\rangle
&=&
\sum_{\bm{m},\bm{l}}K^{(n;\bm{m})}_{j;\bm{l}}
\langle{n_x}'_1{n_x}'_2|\hat{O}^{(m_x)}_{l_x}|{n_x}_1{n_x}_2\rangle
 \nonumber \\ && \times
\langle{n_y}'_1{n_y}'_2|\hat{O}^{(m_y)}_{l_y}|{n_y}_1{n_y}_2\rangle
\langle{n_z}'_1{n_z}'_2|\hat{O}^{(m_z)}_{l_z}|{n_z}_1{n_z}_2\rangle ,
\nonumber \\
\end{eqnarray}
where
\begin{eqnarray}
\label{eq:7}
\langle{n}'_1{n}'_2|\hat{O}^{(m)}_l|{n}_1{n}_2\rangle
&=&
\int\rmd{r}'_1\rmd{r}'_2\rmd{r}_1\rmd{r}_2\,
\delta({r}'_1-{r}_1)\delta({r}'_2-{r}_2)g_a({r}_1-{r}_2)
\nonumber \\ &&\times\,
\hat{O}^{(m)}_l({k}',{k})
\psi_{{n}'_1}({r}'_1)\psi_{{n}'_2}({r}'_2)
\psi_{{n} _1}({r} _1)\psi_{{n} _2}({r} _2),
\end{eqnarray}
and where the 1D Gaussian formfactor (\ref{eq:1a}) reads
$g_a({r})={e^{-{{r}^2}/{a^2}}}/\left(a\sqrt{\pi}\right)$.

The standard way to proceed, which for the local LO term
$\hat{O}^{(0)}_1=1$ was developed in Ref.~\cite{(Gir83a)}, is to replace the products of HO wavefunctions
$\psi_{{n}'_1}({r}_1)\psi_{{n}_1}({r}_1)$ and
$\psi_{{n}'_2}({r}_2)\psi_{{n}_2}({r}_2)$ by sums of the HO wavefunctions.
Then, use the Moshinsky brackets to introduce the relative
coordinate ${r}_1-{r}_2$ on which the Gaussian form factor depends. At higher orders,
this approach requires explicit treatment of terms that stem
from expanding powers of relative-momentum operators that appear in Eq.~(\ref{eq:5}). Although this
tedious procedure was up to 4th order implemented in version {(v\codeversion)}
of the code \pr{hfodd}, an alternative and more
compact procedure is to reverse the order of steps and begin
by performing two Moshinsky transformations,
\begin{eqnarray}
\label{eq:8a}
\psi_{{n}'_1}({r}'_1)\psi_{{n}'_2}({r}'_2) &=& \sum_{N'=0}^{{n}'_1+{n}'_2} M_{N'}^{{n}'_1{n}'_2}\psi_{N'}(R')\psi_{{n}'_1+{n}'_2-N'}({r'}),
\\
\label{eq:8b}
\psi_{{n} _1}({r} _1)\psi_{{n} _2}({r} _2) &=& \sum_{N =0}^{{n} _1+{n} _2} M_{N }^{{n} _1{n} _2}\psi_{N }(R )\psi_{{n} _1+{n} _2-N }({r }),
\end{eqnarray}
for $R'=\tfrac{{r}'_1+{r}'_2}{\sqrt{2}}$,     $r'=\tfrac{{r}'_1-{r}'_2}{\sqrt{2}}$,
    $R =\tfrac{{r} _1+{r} _2}{\sqrt{2}}$, and $r =\tfrac{{r} _1-{r} _2}{\sqrt{2}}$.
Since in Eq.~(\ref{eq:5}), the relative-momentum operators
$k =\frac{-i}{\sqrt{2}}\frac{\partial}{\partial{r }}$ and
$k'=\frac{-i}{\sqrt{2}}\frac{\partial}{\partial{r'}}$
act only on wavefunctions
$\psi_{{n} _1+{n} _2-N }({r })$ and $\psi_{{n}'_1+{n}'_2-N'}({r'})$,
respectively, the
integrals over $R'$ and $r'$ can be performed, which gives
\begin{eqnarray}
\label{eq:10}
\hspace*{-2cm}
\langle{n}'_1{n}'_2|\hat{O}^{(m)}_l|{n}_1{n}_2\rangle
&=&
 \frac{(-1)^{m/2}}{2^{l+m/2}}
\sum_{N'=0}^{{n}'_1+{n}'_2} M_{N'}^{{n}'_1{n}'_2}
\sum_{N =0}^{{n} _1+{n} _2} M_{N }^{{n} _1{n} _2}
\left[\int\rmd{R}\,\psi_{N'}(R )\psi_{N }(R )
\right]
\nonumber \\
\hspace*{-2cm}
&\times&\left[
\sum_{l'=0}^l\newton{l}{l'}
\int\rmd{r}\,
\psi^{(m/2+2l'-l)}_{{n}'_1+{n}'_2-N'}({r })
\psi^{(m/2-2l'+l)}_{{n} _1+{n} _2-N }({r })
g_a\left(\sqrt{2}{r}\right)
\right],
\end{eqnarray}
where superscripts denote derivatives of wavefunctions:
$\psi^{(i)}_n(r)=\frac{\rmd^i}{\rmd{r}^i}\psi_n(r)$.
Orthogonality of wavefunctions $\psi_{N }(R )$
allows for the presentation of the final result as
\begin{eqnarray}
\label{eq:12}
\langle{n}'_1{n}'_2|\hat{O}^{(m)}_l|{n}_1{n}_2\rangle
&=&
 \frac{(-1)^{m/2}}{2^{l+m/2}}
\sum_{N =0}^{\min{({n}'_1+{n}'_2,{n} _1+{n} _2})}
M_{N }^{{n}'_1{n}'_2}
M_{N }^{{n} _1{n} _2}
 \nonumber \\ && \times
\sum_{l'=0}^l\newton{l}{l'}
C^{(m/2+2l'-l,m/2-2l'+l)}_{{n}'_1+{n}'_2-N,{n} _1+{n} _2-N }(a),
\end{eqnarray}
where
\begin{eqnarray}
\label{eq:13}
C^{(i',i)}_{n',n}(a)
&=&
\int\rmd{r}\,
\psi^{(i')}_{n'}({r })
\psi^{(i )}_{n }({r })
g_a\left(\sqrt{2}{r}\right).
\end{eqnarray}

Similarly as it was shown in Ref.~\cite{(Gir83a)}, a relatively simple
analytic expressions can be derived for coefficients
$C^{(i',i)}_{n',n}(a)$. However, these expressions involve
alternating-sign sums of ratios of large factorials and are thus prone
to generating significant numerical
instabilities~\cite{(Egi97b),(Gon21)}. Here we argue that using
such analytical expressions in practical implementations is
not necessary. Indeed, a very simple and extremely stable numerical
derivation based on Gauss-Hermite quadratures is possible; it was already
implemented in the previous version (v2.73y) of \pr{hfodd}~\cite{(Sch17d)}
to treat the Gogny force.

In the context of higher-order finite-range functional generators (\ref{eq:1})
discussed here, the numerical implementation works as follows.
First, we represent derivatives of the HO wavefunctions (I-72) as
\begin{eqnarray}
\label{eq:14}
\psi^{(i)}_n(r)&=&\frac{\rmd^i}{\rmd{r}^i}\psi_n(r)\equiv
b^{i+1/2}H^{(i)}_n(\xi)e^{-\xi^2/2},
\end{eqnarray}
where $\xi=br$ is the position $r$ scaled by the oscillator constant $b=\sqrt{m\omega/\hbar}$
and $H^{(i)}_n(\xi)$ are polynomials of order $n+i$,
which can be easily derived from the standard Hermite
polynomials $H^{(0)}_n(\xi)$ and their first derivatives $H^{(0)'}_n(\xi)$, e.g.,
\begin{eqnarray}
H^{(1)}_n(\xi) &=& H^{(0)'}_n(\xi) -\xi H^{(0)}_n(\xi),
\label{eq:15-1} \\
H^{(2)}_n(\xi) &=&  \left(\xi^2-2n-1\right) H^{(0)}_n(\xi),
\label{eq:15-2} \\
H^{(3)}_n(\xi) &=&  \left(\xi^2-2n-1\right) H^{(0)'}_n(\xi) - \left(\xi^3-(2n+3)\xi\right) H^{(0)}_n(\xi),
\label{eq:15-3} \\
H^{(4)}_n(\xi) &=&  4\xi H^{(0)'}_n(\xi) + \left(\xi^4-(4n+6)\xi^2+4n^2+4n+3\right) H^{(0)}_n(\xi).
\label{eq:15-4}
\end{eqnarray}
This allows one to represent Eq.~(\ref{eq:13}) as
\begin{eqnarray}
\label{eq:16}
C^{(i',i)}_{n',n}(a)
&=&  \frac{b^{i'+i}}{a\sqrt{\pi}}
\int\rmd{\xi}\,
H^{(i')}_{n'}(\xi)
H^{(i )}_{n }(\xi)
e^{-\frac{2+b^2a^2}{b^2a^2}\xi^2}
\nonumber \\
&=&  \frac{b^{i'+i+1}}{\sqrt{\pi}\sqrt{2+b^2a^2}}
\int\rmd{\eta}\,
H^{(i')}_{n'}(\epsilon\eta)
H^{(i )}_{n }(\epsilon\eta)
e^{-\eta^2}
\nonumber \\
&=&  \frac{b^{i'+i+1}}{\sqrt{\pi}\sqrt{2+b^2a^2}}
\sum_{k=1}^{K}W_k
H^{(i')}_{n'}(\epsilon\eta_k)
H^{(i )}_{n }(\epsilon\eta_k),
\end{eqnarray}
where
\begin{eqnarray}
\label{eq:16b}
\epsilon&=&\frac{ba}{\sqrt{2+b^2a^2}}
\end{eqnarray}
and $W_k$ and $\eta_k$ are,
respectively, weights and nodes of the Gauss-Hermite quadrature of
order $K=n'+n+i'+j+1$. For calculations employing the HO basis of up
to $N_0$ quanta in the given Cartesian direction $x$, $y$, or $z$,
and for derivatives up to 4th order, the quadrature of order
$K=2N_0+5$ thus gives the exact result and no accumulation of numerical errors
is expected.

Exactly the same method can be used to evaluate the Moshinsky
coefficients, which in their exact analytical form (VI-63)~\cite{(Dob09g)} also
involve numerically unstable alternating-sign sums of ratios of large factorials.
Indeed, by setting in Eq.~(\ref{eq:8b})  ${r}_1={r}_2\equiv{r}$
and inserting Eq.~(\ref{eq:14}) for $i=0$ we obtain
\begin{eqnarray}
\label{eq:17}
H^{(0)}_{{n}_1}(\xi)
H^{(0)}_{{n}_2}(\xi)
&=& \sum_{N =0}^{{n} _1+{n} _2} M_{N }^{{n} _1{n} _2}
H^{(0)}_{N}(\sqrt{2}\xi)
H^{(0)}_{{n} _1+{n} _2-N }({0}).
\end{eqnarray}
We now can multiply both sides by $\sqrt{2}H^{(0)}_{N'}(\sqrt{2}\xi)e^{-2\xi^2}$,
integrate over $\xi$,
and use the orthogonality condition of the Hermite
polynomials on the right-hand side. This finally gives,
\begin{eqnarray}
\label{eq:18}
\int\rmd\eta\,
H^{(0)}_{{n}_1}\left({\textstyle\frac{\eta}{\sqrt{2}}}\right)
H^{(0)}_{{n}_2}\left({\textstyle\frac{\eta}{\sqrt{2}}}\right)
H^{(0)}_{N}(\eta)e^{-\eta^2}
&=& M_{N }^{{n} _1{n} _2}
H^{(0)}_{{n} _1+{n} _2-N }({0}),
\end{eqnarray}
where $\eta=\sqrt{2}\xi$. This allows one to determine the exact Moshinsky
coefficients through a numerically stable Gauss-Hermite quadrature
of order $K=n_1+n_2+N+1$,
\begin{eqnarray}
\label{eq:19}
M_{N }^{{n} _1{n} _2}
&=&
\left(H^{(0)}_{{n} _1+{n} _2-N }({0})\right)^{-1}
\sum_{k=1}^{K} W_K
H^{(0)}_{{n}_1}\left({\textstyle\frac{\eta_k}{\sqrt{2}}}\right)
H^{(0)}_{{n}_2}\left({\textstyle\frac{\eta_k}{\sqrt{2}}}\right)
H^{(0)}_{N}(\eta_k) .
\end{eqnarray}
Therefore, equations (\ref{eq:16}) and (\ref{eq:19}) give an exact and numerically stable
representation of the 1D matrix elements (\ref{eq:12}) of higher-order
generators. Furthermore, coefficients
$M_{N }^{{n} _1{n} _2}$ and $C^{(i',i)}_{n',n}(a)$ have to be calculated only once
and if needed, stored.

In the special case of local generators discussed in Ref.~\cite{(Dob12b)},
the central pseudopotential (\ref{eq:1}) reduces
for
$W^{(n)}\equiv W^{(n)}_1-W^{(n)}_2$,
$B^{(n)}\equiv B^{(n)}_1-B^{(n)}_2$,
$H^{(n)}\equiv H^{(n)}_1-H^{(n)}_2$, and
$M^{(n)}\equiv M^{(n)}_1-M^{(n)}_2$
to
\begin{eqnarray}
\label{eq:20}
 \mathcal{V}^{\text{loc}}_C(\bm{r}'_1\bm{r}'_2;\bm{r}_1\bm{r}_2) &=& \sum_{n}
   \left(W^{(n)} \hat{1}_\sigma\hat{1}_\tau+B^{(n)} \hat{1}_\tau  \hat{P}^\sigma
        -H^{(n)} \hat{1}_\sigma\hat{P}^\tau-M^{(n)} \hat{P}^\sigma\hat{P}^\tau\right)
\nonumber \\
&&~~~~~~~~~~~~\times \delta(\bm{r}'_1-\bm{r}_1)\delta(\bm{r}'_2-\bm{r}_2)
V^{(n)}(\bm{r}_1-\bm{r}_2),
\end{eqnarray}
where
\begin{eqnarray}
\label{eq:21}
V^{(n)}(\bm{r}) \equiv 2^{-n}\Delta^{n/2}g_a(\bm{r})
\end{eqnarray}
and where $\Delta$ is the standard differential Laplace operator.
Explicitly, this gives:
\begin{eqnarray}
V^{(0)}_n(\bm{r}) &=& g_a(\bm{r}),
\label{eq:22-0} \\
V^{(1)}_n(\bm{r}) &=& \tfrac{1}{a^2}\left(2\left(\tfrac{\bm{r}^2}{a^2}\right)-3\right)g_a(\bm{r}),
\label{eq:22-1} \\
V^{(2)}_n(\bm{r}) &=& \tfrac{1}{a^4}\left(4\left(\tfrac{\bm{r}^4}{a^4}\right)
                                        -20\left(\tfrac{\bm{r}^2}{a^2}\right)+15\right)g_a(\bm{r}),
\label{eq:22-2} \\
V^{(3)}_n(\bm{r}) &=& \tfrac{1}{a^6}\left(8\left(\tfrac{\bm{r}^6}{a^6}\right)
                                        -84\left(\tfrac{\bm{r}^4}{a^4}\right)
                                       +210\left(\tfrac{\bm{r}^2}{a^2}\right)-105\right)g_a(\bm{r}).
\label{eq:22-3}
\end{eqnarray}

Similarly, as for the nonlocal operators above in Eq.~(\ref{eq:4}),
we can now rewrite potentials~(\ref{eq:21}) as sums of terms that are products of
powers of positions $x$, $y$, and $z$. That is,
\begin{eqnarray}
\label{eq:4a}
V^{(n)}(\bm{r})&=&\sum_{\bm{m}}K^{(n;\bm{m})}
x^{m_x}
y^{m_y}
z^{m_z},
\end{eqnarray}
where $K^{(n;\bm{m})}\equiv K^{(n;m_x,m_y,m_z)}$
for $0\leq m_x+m_y+m_z\leq{n}$ are integer coefficients.
The matrix element of two-body potential $V^{(n)}(\bm{r}_1-\bm{r}_2)$ is thus equal to the sum of products
of 1D matrix elements,
\begin{eqnarray}
\label{eq:6a}
\hspace*{-15mm}
\langle\bm{n}'_1\bm{n}'_2|V^{(n)}|\bm{n}_1\bm{n}_2\rangle
= \tfrac{1}{\left(a\sqrt{\pi}\right)^3}
\sum_{\bm{m}}K^{(n;\bm{m})}
&&                    \langle{n'_{x_1}}{n'_{x_2}}|({x}_1-{x}_2)^{m_x}{e^{-{({x}_1-{x}_2)^2}/{a^2}}}|{n_{x_1}}{n_{x_2}}\rangle
\nonumber \\[-3.5mm]
&&\hspace*{-5mm}\times\langle{n'_{y_1}}{n'_{y_2}}|({y}_1-{y}_2)^{m_y}{e^{-{({y}_1-{y}_2)^2}/{a^2}}}|{n_{y_1}}{n_{y_2}}\rangle
\nonumber \\
&&\hspace*{-5mm}\times\langle{n'_{z_1}}{n'_{z_2}}|({z}_1-{z}_2)^{m_z}{e^{-{({z}_1-{z}_2)^2}/{a^2}}}|{n_{z_1}}{n_{z_2}}\rangle,
\end{eqnarray}
where
\begin{eqnarray}
\label{eq:7a}
\langle{n}'_1{n}'_2|({r}_1-{r}_2)^{m}{e^{-{({r}_1-{r}_2)^2}/{a^2}}}|{n}_1{n}_2\rangle
&=&
\int\rmd{r}_1\rmd{r}_2\,
({r}_1-{r}_2)^{m}e^{-{({r}_1-{r}_2)^2}/{a^2}}
\nonumber \\ &&\times\,
\psi_{{n}'_1}({r}_1)\psi_{{n}'_2}({r}_2)
\psi_{{n} _1}({r}_1)\psi_{{n} _2}({r}_2)
\nonumber \\
&\hspace*{-15mm}=&\hspace*{-15mm}
\sum_{N =0}^{\bar{N}}
M_{N }^{{n}'_1{n}'_2}
M_{N }^{{n} _1{n} _2}
C^{(m)}_{{n}'_1+{n}'_2-N,{n} _1+{n} _2-N }(a),
\end{eqnarray}
and where $\bar{N}=\min{({n}'_1+{n}'_2,{n} _1+{n} _2})$ and
\begin{eqnarray}
\label{eq:16a}
C^{(m)}_{n',n}(a)
&=&  \frac{b^{1-m}}{\sqrt{\pi}\sqrt{2+b^2a^2}}
\sum_{k=1}^{K}W_k
H^{(0)}_{n'}(\epsilon\eta_k)
H^{(0)}_{n }(\epsilon\eta_k)
\times      (\epsilon\eta_k)^m
\end{eqnarray}
for $\epsilon$ given in Eq.~(\ref{eq:16b})
and $W_k$ and $\eta_k$ being,
respectively, weights and nodes of the Gauss-Hermite quadrature of
order $K=n'+n+m+1$.

\subsection{Finite-range separable terms}
\label{subsec:Separable}

The separable pairing force in the isovector $^1S_0$ channel,
introduced by Tian {\it et al.} in the spherical case~\cite{(Tia09d)}
and by Nik{\v{s}}i{\'c} {\it et al.} in the 3D
deformed case~\cite{(Nik10),(Nik14)}, is implemented
in version {(v\codeversion)} of the code \pr{hfodd}. The general expression of this
interaction in the 3D Cartesian coordinates is
\begin{eqnarray}
\label{eq:101}
 \mathcal{V}_S(\bm{r}'_1\bm{r}'_2;\bm{r}_1\bm{r}_2) &=&
   \left(\tilde{W} \hat{1}_\sigma\hat{1}_\tau+\tilde{B} \hat{1}_\tau  \hat{P}^\sigma
        -\tilde{H} \hat{1}_\sigma\hat{P}^\tau-\tilde{M} \hat{P}^\sigma\hat{P}^\tau\right) \nonumber \\
&&\times  \delta(\bm{R}'-\bm{R}) P(\bm{r}')P(\bm{r}),
\end{eqnarray}
where formfactor $P(\bm{r})$ is equal to a sum of Gaussians (\ref{eq:1a}),
\begin{eqnarray}
\label{eq:102}
P(\bm{r}) &=& \sum_k^K A_kg_{a_k}(\bm{r}),
\end{eqnarray}
and $\bm{r}'$=$\bm{r}'_1-\bm{r}'_2$, $\bm{r}$=$\bm{r}_1-\bm{r}_2$,
$\bm{R}'$=$\thalf (\bm{r}'_1+\bm{r}'_2)$, and $\bm{R}$=$\thalf (\bm{r}_1+\bm{r}_2)$
are the relative and center-of-mass coordinates. To avoid redundancy
with coupling constants $\tilde{W}$, $\tilde{B}$, $\tilde{H}$, and $\tilde{M}$,
one should use normalisation
$\sum_k A_k=1$.

A detailed derivation of the matrix elements of the separable
generators in zero order can be found in~\cite{(Rom20)} and
references therein. This implementation was compared with an updated
version of the code \pr{hosphe}~\cite{(Car10d)}, where the separable
interaction was implemented in spherical symmetry. The figure of
Ref.~\cite{(Ves11a)} was reproduced up to a precision of 1\,eV,
therefore confirming the accuracy of our implementation.

\subsection{Zero-range two-body pairing terms}
\label{subsec:pairing}

In version {(v\codeversion)} of the code \pr{hfodd}, all terms of the pairing
functional~\cite{(Per04c)} that correspond
to the Skyrme functional were implemented.

\subsection{Multi-quasiparticle blocking}
\label{subsubsec:Multi-particle}

The quasiparticle blocking was initially introduced in version
(v2.40h) of the  \pr{hfodd} code (see VI) to allow for the
description of odd-$A$ or odd-odd paired nuclei.  It consists in
looking for a solution of the HFB equations with as ansatz a vacuum
$|\Phi\rangle$ onto which a single-quasiparticle excitation
$\beta^\dagger_k$ is applied,
$|\Phi_k\rangle=\beta^\dagger_k|\Phi\rangle$. Specifically, at each
iteration, the code selects the quasiparticle state $k$ in the matrix
$\varphi$ and exchanges its upper ($B^*$) and lower ($A^*$)
components with those ($A$, $B$) of its partner of opposite
quasiparticle energy, see Eqs.~(VI-83)--(VI-86)~\cite{(Dob09g)}.
According to the option requested by the user, see keywords \tk{BLOCKFIX\_N}
or \tk{BLOCKFIX\_P} in Section~VI-3.3~\cite{(Dob09g)}, the blocked
quasiparticle $k$ may be kept the same throughout the calculation, or
selected at each iteration as the one having the maximum overlap with
a single-particle state (or its time-reversed image) chosen
beforehand.

In version {(v\codeversion)} of the code \pr{hfodd}, this method was
extended to HFB states $|\Phi_{\bm{k}}\rangle$ with an
arbitrary number $r$ of quasiparticle excitations $\bm{k}\equiv(
k_1,\ldots k_r)$
\begin{equation} \label{multblock}
|\Phi_{\bm{k}}\rangle= \prod_{\nu=1}^r \beta^\dagger_{k_\nu}|\Phi \rangle,
\end{equation}
where $|\Phi\rangle$ is the HFB vacuum for the quasiparticle
operators $\beta_k$, $k=1,\ldots,M$. The wave function in
Eq.~(\ref{multblock}) is represented by the $2M\times M$ matrices
$(\varphi_{\bm{k}},\chi_{\bm{k}})$ obtained by swapping the
components of the blocked quasiparticles in the solutions of the HFB
solution $(\varphi,\chi)$ . For instance, the wave functions
associated with negative energies may schematically be written as
\begin{equation}
   \varphi_{\bm{k}} = \left(\begin{array}{cccccccc}
   B^*_1 & \dots & A_{k_1} &  B^*_{k_1+1} & \dots &  A_{k_r} & \dots & B^*_M  \\
   A^*_1 & \dots & B_{k_1} &  A^*_{k_1+1} & \dots &  B_{k_r} & \dots & A^*_M
   \end{array}\right).
\end{equation}
Numerically, the quasiparticle $k_\nu$, $\nu=1\rightarrow r$, are
defined \textit{via} successive applications of the procedure
described above, whilst ensuring a given label can be selected only
once.

\subsection{Pfaffian overlaps}
\label{subsubsec:Pfaffian}

Computation of the overlap kernels between HFB wave functions is of
crucial importance in multi-reference calculations and for symmetry
restoration.  Up to recently, such scalar products were most often
evaluated using the Onishi formula \cite{(Oni66)}, which suffers from a
sign ambiguity due to a square-root appearing there. This limitation was then
overcome by Robledo \textit{via} a new expression involving a Pfaffian
\cite{(Rob09)}. Version {(v\codeversion)} of the code \pr{hfodd} includes a new module,
based on an equivalent Pfaffian formulation derived in Ref.~\cite{(Ber12b)},
which allows for determining the overlap between two arbitrary, potentially
blocked HFB states of the general form (\ref{multblock}) as
\begin{equation} \label{HFBPfaf}\hspace*{-22mm}
\langle \Phi_{\bm{k}} | \Phi'_{\bm{k}'} \rangle = (-1)^{M(M-1)/2} (-1)^{r(r-1)/2} \mathrm{pf}
\left(\begin{array}{cccc}
B^TA               &  B^T p^\dagger                &  B^T q'^T             & B^TB'^*   \\
-p^*B              &  q^* p^\dagger                &  q^* q'^T               & q^* B'^*   \\
-q'B                 &   -q' q^\dagger                  &   p' q'^T                 &   p' B'^*   \\
-B'^\dagger B &  - B'^\dagger q^\dagger   & -B'^\dagger p'^T   &   A'^\dagger B'^*
\end{array}\right).
\end{equation}
This relation holds for non-normalized wave functions. Square matrices
$A$ and $B$ ($A'$ and $B'$) stand for the usual blocks of the Bogolyubov
transformation, which correspond to the non-blocked HFB state $|\Phi \rangle$ ($|\Phi' \rangle$),
whereas rectangular matrices $p$ and $q$ ($p'$ and $q'$) contain
components of the $r$ ($r'$) blocked quasiparticle states,
see Ref.~\cite{(Ber12b)}. In the
case of single-quasiparticle blocking, the latter matrices reduce to row vectors, whereas
for non-blocked HFB wave functions, the corresponding rows and columns do not appear in
matrix (\ref{HFBPfaf}).

Equation (\ref{HFBPfaf}) is valid only for a complete quasiparticle
space. Consequently, the Pfaffian formula cannot be used when a cut-off
in the space of quasiparticle states is implemented, see Section~IV-3.1~\cite{(Dob04d)}.
Therefore, to use the Pfaffian formula, the
pairing cut-off must be handled within the two-basis method, see Section~VII-2.2.1~\cite{(Sch12c)}.
Moreover, since the Pfaffian formula is based on associating the non-blocked (even) HFB state
with the product of all quasiparticle annihilation operators acting on the
true vacuum, $|\Phi \rangle\propto\prod_i\beta_i|0\rangle$,
the number of quasiparticles must be even. This implies that the dimension of the single-particle
space generated by the two-basis method must be even.

\subsection{Particle-number and parity symmetry restoration}
\label{subsubsec:Symmetry}

The HFB method accounts for pairing correlations through the breaking
of the U(1) symmetry associated with particle-number conservation.
The above computation of overlaps allows us to implement the
restoration of correct proton and neutron numbers by projection after
variation of a symmetry-unrestricted HFB state
$|\Phi\rangle$~\cite{(She19c)}. In version {(v\codeversion)} of the
code \pr{hfodd}, projections on the total particle number $A$ and
isospin projection $T_z=(N-Z)/2$ were implemented by introducing
two new independent keywords.
Activating only one of those projections thus allows for a full particle-number-symmetry
restoration for nuclei where one of the species, protons or
neutrons, are unpaired in $|\Phi\rangle$.
In future releases, this implementation will be
optimised by considering a 1D gauge-angle integration that allows
for the simultaneous restoration of both proton- and neutron-number
symmetries, according to the methodology presented in
Ref.~\cite{(Ang01)} and routinely used in other implementations,
see, e.g., Ref.~\cite{(Bal21)}.

Version {(v\codeversion)} of the code \pr{hfodd} also
incorporates the parity restoration by means of the projector
$\hat{P}_\pi = (1+\pi\hat{\Pi})/2$ where $\pi=\pm$ and $\hat{\Pi}$ is the
inversion transformation. Finally, a state with good quantum
numbers $A$, $T_z$, $I^\pi$, $M$, and $K$ is obtained as
\begin{eqnarray}  \label{projHFB}
\hspace*{-2.5cm}
|AT_z;I^\pi MK \rangle &=& \hat{P}_A \hat{P}_{T_z}  \hat{P}_{MK}^I \hat{P}_{\pi} |\Phi \rangle \\
&=& \frac{2I+1}{16\pi^4} \!\! \int_0^{\pi}  \!\!  \!\! d\phi e^{-i\phi A}
 \!\! \int_0^{\pi} \!\! \!\!   d\phi_T e^{-i \phi_T T_z} \!\! \int \!\!  d\Omega D^{I*}_{MK} (\Omega)
 e^{i\phi \hat{A}} e^{i\phi_T \hat{T}_z}  \hat{R}(\Omega) (1+\pi\hat{\Pi})|\Phi \rangle \nonumber  .
\end{eqnarray}
The operations detailed in Section VI-2.1~\cite{(Dob09g)} for angular-momentum
projection of Slater determinants were generalized to the HFB states.
Quasiparticles $\tilde{\varphi}^T$, transformed by  generic symmetry
operators $\hat{T}$ that appear in Eq.~(\ref{projHFB}), read
\begin{eqnarray}  \label{eq:quasi}
\tilde{\varphi}^T&=&\left(\begin{array}{cc}T & 0 \\0 & T^* \end{array}\right)
                    \left(\begin{array}{c} B^*\\   A^*\end{array}\right)
                  = \left(\begin{array}{c}TB^*\\T^*A^*\end{array}\right),\quad\mbox{for}\quad
\tilde{\varphi}   = \left(\begin{array}{c} B^*\\   A^*\end{array}\right)
\end{eqnarray}
characterizing state $|\Phi\rangle$ and $T$ denoting
representation of $\hat{T}$ in the single-particle
basis. Then, kernels of observables are computed according to the
generalised Wick's theorem~\cite{(Rin80b)} in terms of the transition normal and pairing
densities, and the overlap kernels are evaluated according to Eq.~(\ref{HFBPfaf}).

In version {(v\codeversion)} of the
code \pr{hfodd}, particle-number projection is realized by using the
Gauss-Tchebyschev quadratures, whereas the discrete parity projector is
applied explicitly. The numerical treatment of the integration over the
Euler angles $\Omega$ was described in Section VI-2.1~\cite{(Dob09g)}.

\subsection{Axialization}
\label{subsec:Axialization}

In version {(v\codeversion)} of the code \pr{hfodd}, axial
self-consistent solutions can be obtained by projecting wave
functions on the axial shape, with the symmetry
axis oriented along the $z$ axis. This is achieved by projecting the
particle-hole or pairing mean-field and/or particle-hole or pairing
density matrix (pairing tensor) on those corresponding to the axial
symmetry. Specifically, this is achieved by expanding the Cartesian
harmonic-oscillator basis used by the code on states having good
quantum numbers $\Omega_k$, which are the eigenvalues of the $z$
component of the single-particle angular momentum. Then, at each
iteration, only the matrix elements of the particle-hole matrices
that are diagonal in $\Omega_k$ are kept and/or only the off-diagonal
($\Omega_k,-\Omega_k$) matrix elements of the pairing matrices are kept.
At convergence, an axial state is obtained with all single-particle
or quasiparticle states having good quantum numbers $\Omega_k$.
The axialization helps to stabilizes the convergence of states, which
at self-consistency are axial, but during the convergence
can wander towards non-axial deformations and thus converge
slowly or sometimes never.

\subsection{Wigner functions}
\label{subsec:Wigner}

To perform the angular-momentum- and isospin-projection calculations,
previously the code \pr{hfodd} used the Wigner formula to compute the
Wigner $d$ functions, $d_{m,n}^j(\theta)$. For $j\ge50$, the Wigner
formula is known to suffer from a loss of precision~\cite{(Taj15)},
which is due to the fact that with $j>>1$ and $\theta\ne0,\pi$, it
relies on a cancellation of very large terms with alternating
signs.

In Ref.~\cite{(Taj15)}, a robust procedure of computing the $d$
functions was proposed. In this method, the $d$ functions were
expended using the Fourier series. In Ref.~\cite{(Fen15)}, another
method was proposed, which was based on the diagonalization of the
angular-momentum operator $J_y$ in the basis of eigenstates of $J_z$.
Version {(v\codeversion)} of the code \pr{hfodd}, after implementing
a few corrections, uses the code published in Ref.~\cite{(Fen15)}.

\subsection{Choice of the harmonic-oscillator basis}
\label{sec:basis}

To fix the HO basis used in the program (see Section II-4~\cite{(Dob97c)})
one needs to choose suitable oscillator constants $b_k$ or, equivalently,
oscillator frequencies $\omega_k$ or oscillator lengths, $L_k$, in three
Cartesian directions,
\begin{equation}
b_k=\sqrt{\frac{m\omega_k}{\hbar}},\quad
L_k=\sqrt{\frac{\hbar}{m\omega_k}},\quad\mbox{for}\quad  k=x,y,z.
\end{equation}
There are many possible ways to determine $b_k$ or $\omega_k$. The
methods implemented in version {(v\codeversion)} of the code
\pr{hfodd} are described below. They correspond to different values of
variable \tv{INPOME} set by using new functionalities of keyword \tk{FREQBASIS}, see
Section~\ref{subsubsec:Previous}, which are described below.

\begin{enumerate}

\item[(o)]\label{en:surf} For the default value \tv{INPOME}=0,  the code
uses values of the basis-deformation input parameters $\alpha_{2\mu}$ to
define surface $\Sigma$,
\begin{equation}\label{eq:ogo}
\Sigma: \ \ \ R(\theta,\phi)=c(\alpha)\left(1+\sum_{\lambda\mu}\alpha_{\lambda\mu}Y^{*}_{\lambda\mu}(\theta,\phi)\right),
\end{equation}
and then it determines mean squared values of positions $r_k^2$ over the
interior of $\Sigma$:
\begin{equation}
R^2_k=\int_{r<R(\theta,\phi)} r^2_k dV, \ \ \ k=x,y,z.
\end{equation}
Conditions
\begin{equation}\label{eq:omr}
\omega_xR_x=\omega_yR_y=\omega_zR_z
\end{equation}
and
\begin{equation}\label{eq:om3}
\omega_x\omega_y\omega_z=\omega_0^3
\end{equation}
are then used to determine $\omega_k$, with
$\omega_0$ calculated according to Eq.~(I-3)~\cite{(Dob97b)}.
Parameter $c(\alpha)$ is fixed by the condition that the volume $V_0$ inside
$\Sigma$ is equal to
\begin{equation}
V_0=\tfrac{4}{3}\pi R_0^3,
\end{equation}
where $R_0=r_0A^{1/3}$, and  $r_0$ is given by variable \tv{R0PARM} read
under keyword \tk{SURFACE\_PAR}, see Section~II-3.6~\cite{(Dob97c)}. This
prescription works well assuming that $\alpha_{\lambda\mu}$ are real and
$\alpha_{21}=0$, which means, among others, that surface (\ref{eq:ogo}) is
in the principal-axes frame of the quadrupole deformation.

\item For \tv{INPOME}=1, the oscillator frequencies $\omega_k$ are given
explicitly as input parameters of the program; $\omega_x$=\tv{BASINX},
$\omega_y$=\tv{BASINY}, $\omega_z$=\tv{BASINZ}.

\item For \tv{INPOME}=2, the oscillator lengths $L_k$ are given explicitly
as input parameters of the program; $L_x$=\tv{BASINX}, $L_y$=\tv{BASINY},
$L_z$=\tv{BASINZ}.

\item For \tv{INPOME}=3, the oscillator constants $b_k$ are given explicitly
as input parameters of the program; $b_x$=\tv{BASINX}, $b_y$=\tv{BASINY},
$b_z$=\tv{BASINZ}.

\item For \tv{INPOME}=4, to calculate the oscillator lengths $L_k$ the code uses value of
the basis-deformation input parameter $\alpha_{20}$,
\begin{eqnarray}
&L_x=L_0\exp(-\sqrt{\tfrac{5}{16\pi}}\alpha_{20}),\\
&L_y=L_0\exp(-\sqrt{\tfrac{5}{16\pi}}\alpha_{20}),\\
&L_z=L_0\exp(\sqrt{\tfrac{5}{4\pi}}\alpha_{20}),
\end{eqnarray}
in analogy to Eq.~(1.88) in Ref.~\cite{(Rin80b)}. This prescription generates an axial basis.
Here, the code uses $L_0=\sqrt{2*20.73553/\hbar\omega_0}$ for $\hbar\omega_0=1.2*41*A^{-1/3}$.

\item For \tv{INPOME}=5,  the code uses values of mass quadrupole constraints,
$\bar{Q}_{20}$ and $\bar{Q}_{22}$ (in barn), see keyword \tk{MULTCONSTR}
in  Section~II-3.7~\cite{(Dob97c)},
to calculate $\beta$ and $\gamma$ deformation parameters as:
\begin{eqnarray}
\beta &=&C\sqrt{\bar{Q}_{20}^2 + \bar{Q}_{22}^2},\quad
\gamma = {\rm atan\,}(\bar{Q}_{22}/\bar{Q}_{20}),
\end{eqnarray}
where
\begin{equation}\label{eq:ce}
C=10^2\frac{\sqrt{5\pi}}{3AR^2_0}, \ \ R_0=r_0A^{1/3},
\end{equation}
and  $r_0$ (in fm) is given by variable \tv{R0PARM}
read under keyword \tk{SURFAC\_PAR}, see Section~II-3.6~\cite{(Dob97c)}.
Then, to fix frequencies $\omega_k$, the code employs conditions
(\ref{eq:omr}) and (\ref{eq:om3}) with
\begin{equation}
R_k=R_0\left(1+\sqrt{\tfrac{5}{4\pi}}\beta\cos(\gamma-2k\pi/3)\right).
\end{equation}

\item For \tv{INPOME}=6,  the code uses values of
the basis-deformation input parameters $\alpha_{20}$ and $\alpha_{22}$
to calculate oscillator frequencies $\omega_k$ as
\begin{equation}
\omega_k=\omega_0\exp\left(-\sqrt{\tfrac{5}{4\pi}}\beta\cos(\gamma-2k\pi/3)\right),
\end{equation}
where
\begin{eqnarray}
\beta&=&\sqrt{\alpha_{20}^2 + 2\alpha_{22}^2},\quad
\gamma={\rm atan\,}(\sqrt{2}\alpha_{22}/\alpha_{20}),
\end{eqnarray}
and $\omega_0$ is calculated according to Eq.~(I-3)~\cite{(Dob97b)}.

\item For \tv{INPOME}=7,  the code uses values of mass multipole
constraints, $\bar{Q}_{\lambda\mu}$ (in barn$^{\lambda/2}$), see keyword \tk{MULTCONSTR} in
Section~II-3.7~\cite{(Dob97c)}, to calculate Bohr deformations
$\alpha_{\lambda\mu}$ according to the method presented in Section
VI-2.5~\cite{(Dob09g)}, see also keyword \tk{BOHR\_BETAS} in Section
VI-3.5~\cite{(Dob09g)}. Values of $\alpha_{\lambda\mu}$ are then used
in the algorithm developed for \tv{INPOME}=0, see point~(o) above.
\end{enumerate}

Options  \tv{INPOME}=5 and 7 were developed to automatically adjust
the HO basis to the quadrupole constraints requested in, e.g., fission-barrier
calculations. However, a new functionality of keyword \tk{MULTCONSTR},
see Section~\ref{subsubsec:Previous}, allows for reading values of
$\bar{Q}_{\lambda\mu}$ irrespective of whether they are used as constraints.

Options  \tv{INPOME}=0 and 4--7 ignore values of input data \tv{BASINX}, \tv{BASINY}, and \tv{BASINZ}.

Options \tv{INPOME}=0,4,6 use values of the
basis-deformation input parameters $\alpha_{2\mu}$ read under keyword
\tk{SURFAC\_DEF}, see Section~II-3.6~\cite{(Dob97c)}. However, for
\tv{IBCONT}=1, see keyword \tk{CONT\_BASIS}, values read from the
basis file override those read under keyword \tk{SURFAC\_DEF}.

Note also that in the parallel mode of code \pr{hfodd}, the basis
deformation can be automatically adjusted by setting \tv{IBASIS}=1
under keyword \tk{BASISAUTOM}, see
Section~VIII-3.1.4~\cite{(Sch17d)}.

\subsection{Fixed \boldmath{$\Omega$} partitions}
\label{subsec:partitions}

In version {(v\codeversion)} of the code \pr{hfodd}, without pairing and for broken simplex symmetry,
arbitrary partitions of particles among different $\Omega$ blocks were implemented,
where $\Omega$ denotes the eigenvalue of a given Cartesian
component of the single-particle angular momentum on the axial-symmetry axis.
To this end, every single-particle state with the calculated projection of the angular momentum
equal to $\Omega_i$ is attributed to a given $\Omega$
block if $\Omega-\thalf\leq\Omega_i<\Omega+\thalf$. Although this
attribution can always be performed, it can serve its purpose only if
the single-particle states are eigenstates of the given Cartesian component of
the angular momentum, that is, their alignments are properly
quantized. This requires that (i) the nucleus has an axial shape and
(ii) the Kramers degeneracy is lifted by aligning individual angular
momenta along the symmetry axis. The first requirement can be fulfilled
by constraining the non-axial quadrupole deformation to zero,
see Section~II-3.7~\cite{(Dob97c)}, or better, by using the axialization option described in
Sections~\ref{subsec:Axialization} and~\ref{subsubsec:Configurations}.
The second requirement can be fulfilled by using a small value
($\approx1$\,keV) of the cranking frequency along the symmetry axis.
Indeed, when an unpaired nucleus has the axial symmetry with the
symmetry axis aligned with the given Cartesian direction, cranking
along the that axis does not change the single-particle wave functions, but only splits the
corresponding single-particle energies as required. A soft attribution condition specified
above allows for a correct convergence to an axial state even if
during the convergence one or both requirements (i) and (ii) are only
approximately fulfilled.

\subsection{Consistency formula between energy and fields}
\label{subsec:Consistency}

Many authors of Hartree-Fock solvers have implemented a consistency
formula, which allows one to check, by summing over the energies of the occupied
single particle states, that the total energy and the mean-field are consistent
in their implementation (see for example~\cite{(Cot78),(Rys15)}). Such a formula
was also used to define the stability energy employed in the code \pr{hfodd}
as a criterion to terminate iterations, see Eq.~(I-37)~\cite{(Dob97b)}. Here we show that
the consistency formula can be extended to the case of the HFB
calculations and to the energy density which contains linear, bilinear, trilinear,
quadrilinear, or possibly higher couplings of densities. Note that this energy
density is not necessarily derived from an interaction; it is sufficient that it contains
products of densities that are contractions of $n$ creation and $n$
annihilation operators evaluated in a HFB state $|\Phi\rangle$.

Up to $n=4$, the total energy of a nucleus can be split as
\begin{equation}
E=E_1+E_2+E_3+E_4 \label{eq:magic1}
\end{equation}
with
\begin{equation}
E_1=\sum_{ij} v^{(1)}_{ij}\,\rho_{ji} \,,
\end{equation}
\begin{equation}
E_2=\sum_{ijkl} \left(v^{(2)}_{ij,kl}\,\rho_{ki}\rho_{lj}
+\tilde v^{(2)}_{ij,kl}\,\kappa^*_{ij}\kappa_{kl}
\right) \,,
\end{equation}
\begin{equation}
E_3=\sum_{ijklmn}\left(v^{(3)}_{ijk,lmn}\,\rho_{li}\rho_{mj}\rho_{nk}
+\tilde v^{(3)}_{ijk,lmn}\,\kappa^*_{ij}\kappa_{lm}\rho_{nk}
\right) \,,
\end{equation}
\begin{eqnarray}
E_4=\sum_{ijklmnop}\left(v^{(4)}_{ijkl,mnop}\,\rho_{mi}\rho_{nj}\rho_{ok}\rho_{pl}
\right.&+\tilde v^{(4)}_{ijkl,mnop}\,
\kappa^*_{ij}\kappa_{mn}\rho_{ok}\rho_{pl} \nonumber \\
&\left.+\tilde{\tilde v}^{(4)}_{ijkl,mnop}\,
\kappa^*_{ij}\kappa^*_{kl}\kappa_{mn}\kappa_{op}
\right) \,,
\end{eqnarray}
where $v^{(1)}$ is the one-body kinetic operator and $v^{(2)}$,
$\tilde v^{(2)}$, $v^{(3)}$, $\tilde v^{(3)}$, $v^{(4)}$, $\tilde
v^{(4)}$, and $\tilde{\tilde v}^{(4)}$ are 2-, 3-, and 4-body scalar
hermitian matrix elements, which fulfill the same usual properties
under the exchange of indices as the matrix elements of interactions,
and the standard density matrix and pairing tensor of state
$|\Phi\rangle$ are given by
$\rho_{ij}=\langle\Phi|a^\dagger_ja_i|\Phi\rangle$ and
$\kappa_{ij}=\langle\Phi|a_ja_i|\Phi\rangle$\,.

For simplicity, the consistency formula is here derived assuming one species
of nucleons only, that is, only one chemical potential $\lambda$.
We also do not consider the possibility that the $n$-body matrix elements depend
on the one-body density (as is the case when they are derived from
a density-dependent interaction or when the Slater approximation is used
for the Coulomb-exchange term of the energy). The generalization for
such cases is straightforward.

From the energy~(\ref{eq:magic1}), one obtains the normal field
\begin{eqnarray}
h_{ij}&=v^{(1)}_{ij} \nonumber \\
&+\sum_{kl} 2\,v^{(2)}_{ik,jl}\,\rho_{lk} \nonumber \\
&+\sum_{klmn}\left( 3\,v^{(3)}_{ikl,jmn}\,\rho_{mk}\rho_{nl}
+\tilde v^{(3)}_{lki,mnj}\,\kappa^*_{kl}\kappa_{mn}\right) \nonumber \\
&+\sum_{klmnop}\left(4\,v^{(4)}_{iklm,jnop}\,\rho_{nk}\rho_{ol}\rho_{pm}
+2\,\tilde v^{(4)}_{lkmi,nopj}\,\kappa^*_{lk}\kappa_{no}\rho_{pm}\right)
\end{eqnarray}
and the pairing field
\begin{eqnarray}
\tilde h_{ij}
&=\sum_{kl}2\,\tilde v^{(2)}_{ij,kl}\,\kappa^*_{lk} \nonumber \\
&+\sum_{klmn} 2\,\tilde v^{(3)}_{kln,jim}\,\kappa^*_{kl}\rho_{mn} \nonumber \\
&+\sum_{klmnop}\left(
2\,\tilde v^{(4)}_{klmn,jiop}\,\kappa^*_{kl}\rho_{om}\rho_{pn}
+4\,\tilde{\tilde v}^{(4)}_{klmn,opji}
\,\kappa^*_{kl}\kappa^*_{mn}\kappa_{op}
\right)\,.
\end{eqnarray}
Assuming the HFB equations have been solved, the quasiparticle wave-functions
spinors
\begin{equation}
\Psi_j=\left(\begin{array}{c}
U_j \\ V_j
\end{array}\right)
\end{equation}
fulfill the equations
\begin{eqnarray}
\sum_jh_{ij}U_j + \tilde h_{ij} V_j&=&
\left(E_i+\lambda\right)U_i\,, \\
\sum_j\tilde h^*_{ij}U_j - h^*_{ij} V_j&=&
\left(E_i-\lambda\right)V_i
\end{eqnarray}
where $E_i$ are the (positive) quasiparticle energies.
Multiplying the second equation by $V_i^*$ and summing over $i$, one obtains the consistency formula:
\begin{equation}
\label{eq:magic2}
E_1+2E_2+3E_3+4E_4=\sum_iV_i^2\left(\lambda-E_i\right)\,.
\end{equation}
This allows one to define the HFB stability energy,
\begin{equation}
\label{eq:magic}
\delta{\cal E}_{\text{HFB}} = \sum_iV_i^2\left(\lambda-E_i\right) - (E_1+2E_2+3E_3+4E_4)\,,
\end{equation}
which can be used as a measure of deviation of state $|\Phi\rangle$
from the self-consistent solution.

\subsection{Corrected errors}
\label{subsec:bugs}

In version {(v\codeversion)} of the code \pr{hfodd}, we corrected a few
little significant  errors and two significant  errors, Sections~\ref{subsubsec:Yukawa}
and~\ref{subsubsec:gcm},
found in the previous versions of \pr{hfodd}.

\subsubsection{Incorrect signs of the Yukawa energies.}
\label{subsubsec:Yukawa}

In the published versions (v2.08i)~\cite{(Dob04d)},
(v2.08k)~\cite{(Dob05h)}, (v2.40h)~\cite{(Dob09g)}, (v2.49t)~\cite{(Sch12c)}, and (v2.73y)~\cite{(Sch17d)}
of code \pr{hfodd}, signs of the Yukawa energies were inverted. This error was corrected
in the results published in Ref.~\cite{(Dob18a)}.

\subsubsection{Incorrect off-diagonal generator coordinate method (GCM) kernels.}
\label{subsubsec:gcm}

Between versions (v2.10a) and (v2.99u), calculations
of the off-diagonal GCM kernels were incorrect.
The error manifested itself only when the single-particle
wave functions were not real, and was present in
the published versions (v2.40h)~\cite{(Dob09g)}, (v2.49t)~\cite{(Sch12c)},
and (v2.73y)~\cite{(Sch17d)}.

\subsubsection{Definition of the Schiff moment.}
\label{subsubsec:Schiff}

Between versions (v2.19n) and (v2.80m), the factor of
1/10 usually included in the definition of the standard Schiff moment,
cf.~Eq.~(2) in Ref.~\cite{(Dob18a)},
was missing from the values printed on the output file.
This inconsistent definition was implemented in
the published versions (v2.40h)~\cite{(Dob09g)}, (v2.49t)~\cite{(Sch12c)},
and (v2.73y)~\cite{(Sch17d)}.

\subsubsection{Time-odd symmetries in angular-momentum projection.}
\label{subsubsec:AMP}

Before version (v2.66b), for conserved time-odd symmetries (\tv{ISIMTX}=1,
or \tv{ISIMTY}=1 or \tv{ISIMTZ}=1, see Section~IV-3.2~\cite{(Dob04d)}), the
angular-momentum projection was allowed and might give inconsistent
results. This error was thus present in the published version
(v2.49t)~\cite{(Sch12c)} of \pr{hfodd} and corrected in the published
version (v2.73y)~\cite{(Sch17d)}, however, in the latter publication
it was not described.

\subsubsection{Very large harmonic-oscillator bases.}
\label{subsubsec:HO}

As it turns out, for very large harmonic-oscillator bases of
NOSCIL>36, see Section~II-3.6~\cite{(Dob97c)}, the code may behave
erratically. Therefore, beginning with version (v2.81b), calculations
with  NOSCIL>36 are not allowed. This issue awaits future debugging.

\subsubsection{Inconsistent input data in angular-momentum and isospin projection.}
\label{subsubsec:ISOSTZ}

When keywords \tk{PROJECTGCM} (Section~VI-3.2~\cite{(Dob09g)}) and \tk{PROJECTISO}
(Section~VII-3.1~\cite{(Sch12c)}) were simultaneously used
in the input data file, the type of calculation performed could depend on the order
in which they were used. This contradicted the rules of building
the input data file defined in Section~II-3~\cite{(Dob97c)}.
Moreover, for \tv{IPRROT}=0 (see Section~\ref{subsubsec:Previous}),
the remaining input data read under keyword \tk{PROJECTGCM}
were not ignored, which could trigger the AMP
against the user's intentions.

In version {(v\codeversion)} of the code \pr{hfodd}, variables
\begin{itemize}
\item \tv{IPRROT}  (keyword \tk{PROJECTGCM}),
\item \tv{IPRISO}  (keyword \tk{PROJECTISO}),
\item \tv{IPRNUM}  (keyword \tk{PROJPARNUM}),
\item \tv{IPRVEC}  (keyword \tk{PROJVECNUM}),
\item \tv{IPRPTY}  (keyword \tk{PROJPARITY}),
\end{itemize}
must be synchronized, that is, their non-zero values must all be
equal one to another. Internally, they are replaced by the single
variable \tv{IPRGCM}. For any of these variables equal to 0 (not
equal to 0), the remaining input data read under the corresponding
keyword are ignored (used for defining the corresponding projection).

\subsubsection{Inconsistent input data in tilted angular momentum.}
\label{subsubsec:tilted}

When keywords \tk{OMEGA\_XYZ} and \tk{OMEGA\_RTP}
(Section~IV-3.5~\cite{(Dob04d)}) were simultaneously used
in the input data file, the type of calculation performed could depend on the order
in which they were used. This contradicted the rules of building
the input data file defined in Section~II-3~\cite{(Dob97c)}.
In version {(v\codeversion)} of the code \pr{hfodd}, a simultaneous
use of these two keywords is not any more allowed.

\subsubsection{Incorrect information stored on the kernel file.}
\label{subsubsec:kernel}

For runs without isospin-symmetry restoration, not all kernels were
stored on the kernel file, see Section~VI-3.2~\cite{(Dob09g)}, and
nevertheless those not stored were later used in the printouts.
This was causing differences between results printed in the runs
where the kernels  were  calculated  and  those where the kernels
were read from the kernel file.

\subsubsection{Incorrect information stored on the RECORD file.}
\label{subsubsec:record}

After version (v1.78), Fermi energies, pairing gaps, and Lipkin-Nogami
parameters were incorrectly stored on the RECORD file. As a result, a
smooth continuation of runs with pairing could have been impeded. The
error had no effect on final converged results. It was present in all
published versions of \pr{hfodd} after the pairing was introduced in version
(v2.08i)~\cite{(Dob04d)}.

\subsubsection{Incorrect description of keyword \tk{FILSIG\_NEU}.}
\label{subsubsec:FILSIG}

In version (v2.40h)~\cite{(Dob09g)}, description of keyword \tk{FILSIG\_NEU}
was incorrect. It should have referred  to
{\em{twice}} numbers of particles, that is, it should have read:
"matrices \tv{KOFILG} contain twice  numbers  of  particles
put into the states  between  \tv{KHFILG}  and  \tv{KPFILG},  by
using  for  them   partial   occupation   factors   of
\tv{KOFILG}/(\tv{KPFILG}-\tv{KHFILG}+1)/2".

\section{Input Data File}
\label{sec:input_file}

The rules of building
the input data file were defined in Section~II-3~\cite{(Dob97c)}; in version {(v\codeversion)} of the
code \pr{hfodd} these rules remain exactly the same. All previous
items (keywords) of the input data file remain valid, and several new ones were
added, as described in
Sections~\ref{subsubsec:Interaction}--\ref{subsubsec:Starting}. For
some previous items, new features or new values of variables were
added (Section~\ref{subsubsec:Previous}).

For every keyword listed below, we give the default values and names
of the variables read. Apart from character variables, which must
start at the 13th column of the input line, all other variables are
read in the FORTRAN free format. Nevertheless, it is good practice to
include in the input file the integer or real constants when reading
the INTEGER TYPE [IMPLICIT INTEGER (I-N)] or REAL TYPE [IMPLICIT REAL
(A-H,O-Z)] variables, respectively.

\subsection{Interaction}
\label{subsubsec:Interaction}

\subsubsection{Zero-range central terms. \\[2ex]\hspace*{-1em}}
\label{subsubsec:Zero-range}

\key{2BODYDELTA} 0., 0 = \tv{TWOINP}, \tv{ITWOIN}

\keyspace

\noindent For \tv{ITWOIN}=1, the value of a two-body zero-range
          parameter \tv{TWOINP} is added to the Skyrme parameter
          $t_0$. This option is introduced only for convenience of
          handling the input data in cases when a two-body zero-range
          interaction is handled independently of the Skyrme force.
          For \tv{ITWOIN}=0, the value of \tv{TWOINP} is ignored.

\key{3BODYDELTA} 0., 0 = \tv{THRINP}, \tv{ITHRIN}

\keyspace

\noindent For \tv{ITHRIN}=1, the value of \tv{THRINP} defines
          the three-body zero-range parameter $u_0$,
          Eq.~(\ref{eq:3body:gradientsless}). For \tv{ITHRIN}=0, the
          value of \tv{THRINP} is ignored and the three-body
          zero-range force is not taken into account.

\key{4BODYDELTA} 0., 0 = \tv{FOUINP}, \tv{IFOUIN}

\keyspace

\noindent For \tv{IFOUIN}=1, the value of \tv{FOUINP} defines
          the four-body zero-range parameter $v_0$, Eq.~(\ref{eq:4body:gradientsless}).
          For \tv{IFOUIN}=0, the value of \tv{FOUINP} is ignored and the four-body
          zero-range force is not taken into account.

\key{SKYRMEINPU} 0., 0., 0., 0., 0., 0., 0., 0., 0., 1. = \begin{tabular}[t]{l}
                                                          \tv{T0\_DAT}, \tv{X0\_DAT}, \\
                                                          \tv{T1\_DAT}, \tv{X1\_DAT}, \\
                                                          \tv{T2\_DAT}, \tv{X2\_DAT}, \\
                                                          \tv{T3\_DAT}, \tv{X3\_DAT}, \\
                                                          \tv{WW\_DAT}, \tv{PO\_DAT}
                                                          \end{tabular}
\keyspace

\key{SKYRME\_ERR} 0., 0., 0., 0., 0., 0., 0., 0., 0., 1. = \begin{tabular}[t]{l}
                                                          \tv{T0\_ERR}, \tv{X0\_ERR}, \\
                                                          \tv{T1\_ERR}, \tv{X1\_ERR}, \\
                                                          \tv{T2\_ERR}, \tv{X2\_ERR}, \\
                                                          \tv{T3\_ERR}, \tv{X3\_ERR}, \\
                                                          \tv{WW\_ERR}, \tv{PO\_ERR}
                                                          \end{tabular}

\keyspace

\key{SKYRME\_FAC} 0., 0., 0., 0., 0., 0., 0., 0., 0., 1. = \begin{tabular}[t]{l}
                                                          \tv{T0\_FAC}, \tv{X0\_FAC}, \\
                                                          \tv{T1\_FAC}, \tv{X1\_FAC}, \\
                                                          \tv{T2\_FAC}, \tv{X2\_FAC}, \\
                                                          \tv{T3\_FAC}, \tv{X3\_FAC}, \\
                                                          \tv{WW\_FAC}, \tv{PO\_FAC}
                                                          \end{tabular}

\keyspace

\noindent For the Skyrme-force acronym,
          Section~IV-3.1~\cite{(Dob04d)}), \tv{SKYRME}=INPU, values of
          ten input parameters above correspond to the standard
          Skyrme parameters, $t_0$, $x_0$, $t_1$, $x_1$, $t_2$, $x_2$,
          $t_3$, $x_3$,  $W_0$, and $\alpha$, where $\alpha$ is the
          power of density in the density-dependent term. Each
          parameter is determined as, e.g.,
          $t_0$=\tv{T0\_DAT}+\tv{T0\_ERR}*\tv{T0\_FAC}. The formula
          allows for a systematic modification of the central value
          \tv{T0\_DAT}, shifted by a step \tv{T0\_ERR} multiplied by
          a factor \tv{T0\_FAC}. This is useful when building the
          Jacobian matrix~\cite{(Dob14b)} of derivatives of
          observables over the Skyrme parameters.

\key{SKYRMEPAIR} 0 = \tv{KETAPA}

\keyspace

\noindent For \tv{KETAPA}=1 or 2, the pairing terms of the Skyrme
          functional~\cite{(Dob84a),(Per04c)} are taken into account.
          However, for \tv{KETAPA}=2, the pairing terms generated by
          the spin-orbit force $W_0$ are neglected. For
          \tv{KETAPA}=0, all pairing terms of the Skyrme functional
          are neglected. \tv{KETAPA}=1 allows for a fully
          self-consistent pairing calculations performed for the
          \tv{SKYRME}=SKP Skyrme parameters~\cite{(Dob84a)}, but of
          course it can also be used for any other variant of the Skyrme force.
          \tv{KETAPA}>0 requires \tv{NOZEPA}=0. In version
          {(v\codeversion)} of the code \pr{hfodd}, \tv{KETAPA}>0
          still requires \tv{ISIMPY}=0, \tv{ISIQTY}=0, \tv{IPNMIX}=0,
          \tv{IFTEMP}=0, and \tv{KETA\_T}=0.

\subsubsection{Zero-range three-body gradient terms. \\[2ex]\hspace*{-1em}}
\label{subsubsec:gradient}

\key{3BODYGRAD} \hspace*{-15mm}
                0., 0., 0., 0., 0., 0 = \tv{TGRA10}, \tv{TGRA11}, \tv{TGRA20},
                                        \tv{TGRA21}, \tv{TGRA22}, \tv{IGRAIN}

\keyspace

\noindent For \tv{IGRAIN}=1, the parameters of the three-body
          gradient force (\ref{eq:3body:gradients}) are defined
          as $u_1$=\tv{TGRA10}, $y_1$=\tv{TGRA11}, $u_2$=\tv{TGRA20},
          $y_{21}$=\tv{TGRA21}, and $y_{22}$=\tv{TGRA22}. For
          \tv{IGRAIN}=0, the values of \tv{TGRA10}, \tv{TGRA11},
          \tv{TGRA20}, \tv{TGRA21}, and \tv{TGRA22} are ignored and
          the three-body gradient force is not taken into account.
          In a given run of the code \pr{hfodd}, keyword
          \tk{3BODYGRAD} mast not be simultaneously used with keyword
          \tk{3BODYGRUY}.

\key{3BODYGRUY} \hspace*{-15mm}
                0., 0., 0., 0., 0., 0 = \tv{TGRA10}, \tv{TU1\_Y1}, \tv{TGRA20},
                                        \tv{TU2Y21}, \tv{TU2Y22},  \tv{IGRAIN}

\keyspace

\noindent For \tv{IGRAIN}=1, parameters of the three-body gradient
          force (\ref{eq:3body:gradients}) are defined as
          $u_1$=\tv{TGRA10}, $u_1y_1$=\tv{TU1\_Y1}, $u_2$=\tv{TGRA20},
          $u_2y_{21}$=\tv{TU2Y21}, and $u_2y_{22}$=\tv{TU2Y22}. For
          \tv{IGRAIN}=0, the values of \tv{TGRA10}, \tv{TU1\_Y1},
          \tv{TGRA20}, \tv{TU2Y21}, and \tv{TU2Y22} are ignored and
          the three-body gradient force is not taken into account.
          For keyword \tk{3BODYGRUY}, values of \tv{TGRA10}$=0$ or
          \tv{TGRA20}$=0$ are not allowed. In a given run of the
          code \pr{hfodd}, keyword \tk{3BODYGRUY} must not be
          simultaneously used with keyword \tk{3BODYGRAD}.

\subsubsection{Zero-range tensor terms. \\[2ex]\hspace*{-1em}}
\label{subsubsec:Tensor}

\key{SKYRMETENS} 0., 0., 0 = \tv{TEINPU}, \tv{TOINPU}, \tv{KETA\_T}

\keyspace

\noindent For \tv{KETA\_T}=2, parameters of the zero-range tensor
          force, Eq.~(\ref{eq:tensor}), are defined as
          $t_e$=\tv{TEINPU}, $t_o$=\tv{TOINPU}. For \tv{KETA\_T}=1,
          the values of $t_e$ and $t_o$ correspond to those pre-defined
          for a given Skyrme force selected by its acronym, see
          Section~IV-3.1~\cite{(Dob04d)}. For \tv{KETA\_T}=0, the values
          of \tv{TEINPU} and \tv{TOINPU}, are ignored and the tensor
          force is not taken into account. In version
          {(v\codeversion)} of the code \pr{hfodd}, \tv{KETA\_T}>0
          still requires \tv{IPNMIX}=0 and \tv{KETAPA}=0.

\key{TEN\_ADD\_PM}  0., 0., 0., 0., 0., 0., =  \begin{tabular}[t]{l}
                                               \tv{ASCT\_P},\tv{ASCT\_M}, \\
                                               \tv{AKIT\_P},\tv{AKIT\_M}, \\
                                               \tv{ASPT\_P},\tv{ASPT\_M}
                                               \end{tabular}

\keyspace

\noindent By using keyword \tk{TEN\_ADD\_PM}, tensor coupling
          $B_t^{\rm X}$, $B_t^{\rm F}$, and $B_t^{\nabla s}$ for
          $t=0,1$, see Section~\ref{subsec:tensor}, can be shifted by
          adding values of \tv{ASCT\_X}, \tv{AKIT\_X}, and
          \tv{ASPT\_X} for \tv{X=P,M}, respectively.

\key{TEN\_ADD\_TS}  0., 0., 0., 0., 0., 0., =  \begin{tabular}[t]{l}
                                               \tv{ASCT\_T},\tv{ASCT\_S}, \\
                                               \tv{AKIT\_T},\tv{AKIT\_S}, \\
                                               \tv{ASPT\_T},\tv{ASPT\_S}
                                               \end{tabular}

\keyspace

\noindent Same as for keyword \tk{TEN\_ADD\_PM} but for the tensor
          coupling constants in the total-sum representation, see
          Eqs.~(I-14)--(I-15)~\cite{(Dob97b)} and
          Section~II-3.2~\cite{(Dob97c)}.

\key{TEN\_SCA\_PM}  0., 0., 0., 0., 0., 0., =  \begin{tabular}[t]{l}
                                               \tv{SSCT\_P},\tv{SSCT\_M}, \\
                                               \tv{SKIT\_P},\tv{SKIT\_M}, \\
                                               \tv{SSPT\_P},\tv{SSPT\_M}
                                               \end{tabular}

\keyspace

\noindent By using keyword \tk{TEN\_SCA\_PM}, tensor coupling
          $B_t^{\rm X}$, $B_t^{\rm F}$, and $B_t^{\nabla s}$ for
          $t=0,1$, see Section~\ref{subsec:tensor}, can be scaled by
          multiplying them with values of \tv{SSCT\_X}, \tv{SKIT\_X}, and
          \tv{SSPT\_X} for \tv{X=P,M}, respectively.

\key{TEN\_SCA\_TS}  0., 0., 0., 0., 0., 0., =  \begin{tabular}[t]{l}
                                               \tv{SSCT\_T},\tv{SSCT\_S}, \\
                                               \tv{SKIT\_T},\tv{SKIT\_S}, \\
                                               \tv{SSPT\_T},\tv{SSPT\_S}
                                               \end{tabular}

\keyspace

\noindent Same as for keyword \tk{TEN\_SCA\_PM} but for the tensor
          coupling constants in the total-sum representation, see
          Eqs.~(I-14)--(I-15)~\cite{(Dob97b)} and
          Section~II-3.2~\cite{(Dob97c)}.

\subsubsection{Zero-range isospin-breaking terms. \\[2ex]\hspace*{-1em}}
\label{subsubsec:Isospin}

\key{CBR\_CC\_CL2} 0, 0., 0., 0., 0., 0., 0. = \tv{I\_2CBR},
                                               \begin{tabular}[t]{l}
                                               \tv{T02CBR}, \tv{X02CBR}, \\
                                               \tv{T12CBR}, \tv{X12CBR}, \\
                                               \tv{T22CBR}, \tv{X22CBR}
                                               \end{tabular}

\keyspace

\noindent For \tv{I\_2CBR}=1, class-II ISB terms are included in the calculation with parameters:
          $t_0^{\rm{II}}$=\tv{T02CBR}, $x_0^{\rm{II}}$=\tv{X02CBR},
          $t_1^{\rm{II}}$=\tv{T12CBR}, $x_1^{\rm{II}}$=\tv{X12CBR},
          $t_2^{\rm{II}}$=\tv{T22CBR}, $x_2^{\rm{II}}$=\tv{X22CBR}, see Eq.~(\ref{eq:classII_NLO}).
          Note, that the interaction of class II requires p-n mixing (\tv{IPNMIX}=1).
          In version {(v\codeversion)} of the code \pr{hfodd}, \tv{I\_2CBR}=1
          still requires \tv{IPRGCM}=0.

\key{CBR\_CC\_CL3}  0, 0., 0., 0., 0., 0., 0. = \tv{I\_3CBR},
                                                \begin{tabular}[t]{l}
                                                \tv{T03CBR}, \tv{X03CBR}, \\
                                                \tv{T13CBR}, \tv{X13CBR}, \\
                                                \tv{T23CBR}, \tv{X23CBR}
                                                \end{tabular}

\keyspace

\noindent For \tv{I\_3CBR}=1, class-III ISB terms are included in the calculation with parameters:
          $t_0^{\rm{III}}$=\tv{T03CBR}, $x_0^{\rm{III}}$=\tv{X03CBR},
          $t_1^{\rm{III}}$=\tv{T13CBR}, $x_1^{\rm{III}}$=\tv{X13CBR},
          $t_2^{\rm{III}}$=\tv{T23CBR}, $x_2^{\rm{III}}$=\tv{X23CBR}, see Eq.~(\ref{eq:classIII_NLO}).
          In version {(v\codeversion)} of the code \pr{hfodd}, \tv{I\_3CBR}=1
          still requires \tv{IPRGCM}=0.

\subsubsection{Higher-order regularized terms. \\[2ex]\hspace*{-1em}}
\label{subsubsec:Regularized}

\key{REGULFORCE}  0 = \tv{I\_REGA}

\keyspace

\noindent For \tv{I\_REGA}>0, the average mean-field energies of the
          finite-range regularized central pseudopotentials
          (\ref{eq:1}) or (\ref{eq:20}) are calculated. For
          \tv{I\_REGA}=2 or 3, the corresponding direct
          mean fields are included in the self-consistent mean field.
          For \tv{I\_REGA}=2 or 4, the corresponding
          exchange mean fields are included in the self-consistent mean
          field. Altogether, \tv{I\_REGA}=1 demands calculations of
          contributions to energy only, whereas \tv{I\_REGA}=2 demands
          full self-consistent calculations with both direct and
          exchange mean fields included. For \tv{I\_REGA}=0, the
          finite-range regularized central pseudopotential is
          ignored.

          For \tv{I\_REGA}>0 and \tv{IPAHFB}>1, the code issues a
          warning to the effect that, unless the zero-range pairing
          strengths are explicitly set to zero, see
          Section~IV-3.1~\cite{(Dob04d)}, the corresponding pairing
          still will be active. For \tv{NOZEPA}=1, see
          Section~\ref{subsubsec:Miscellaneous}, the zero-range
          pairing is neglected, and the warning is not printed. In
          version {(v\codeversion)} of the code \pr{hfodd},
          \tv{I\_REGA}>0 still requires \tv{IPNMIX}=0, \tv{IRENMA}=0,
          and \tv{IBROYD}=0.

\key{REGUL\_PAIR}  0 = \tv{IREGPA}

\keyspace

\noindent For \tv{IREGPA}>0, the average pairing energies of the
          finite-range regularized central pseudopotential
          (\ref{eq:1}) or (\ref{eq:20}) are calculated. For
          \tv{IREGPA}=2, the corresponding pairing fields are
          included in the self-consistent pairing field. Altogether,
          \tv{IREGPA}=1 demands calculations of contributions to the
          pairing energy, whereas \tv{I\_REGA}=2 demands full
          self-consistent calculations with pairing fields included.
          For \tv{I\_REGA}=0, the pairing contribution of the
          finite-range regularized central pseudopotential is
          ignored. \tv{IREGPA}>0 requires \tv{I\_REGA}>0 and
          \tv{IPAHFB}>0. In version {(v\codeversion)} of the code
          \pr{hfodd}, \tv{IREGPA}>0 still requires \tv{IPNMIX}=0,
          \tv{IRENMA}=0, and \tv{IBROYD}=0.

\key{REGUCOUPLI}   \hspace*{-20mm}\begin{tabular}[t]{lcl}
                   100,  0,  1.       &=& \tv{IREREJ}(1),\tv{NREREJ}(1),\tv{REGWID}     \\
                    0.,  0.,  0.,  0. &=& \tv{REJVCC}(1,1),\tv{REJVCC}(1,2),\tv{REJVCC}(1,3),\tv{REJVCC}(1,4) \\
                    0.,  0.,  0.,  0. &=& \tv{REJVCC}(2,1),\tv{REJVCC}(2,2),\tv{REJVCC}(2,3),\tv{REJVCC}(2,4) \\
                    \multicolumn{3}{c}{\dotfill} \\
                    0.,  0.,  0.,  0. &=& \tv{REJVCC}(i,1),\tv{REJVCC}(i,2),\tv{REJVCC}(i,3),\tv{REJVCC}(i,4) \\
                   \end{tabular}

\keyspace

\noindent After reading the first line, the code reads
          $i$=\tv{IREREJ}(1) lines with
\begin{itemize}
\item four coupling constants $W^{(n)}_j$, $B^{(n)}_j$, $H^{(n)}_j$,
          and $M^{(n)}_j$ per line (for \tv{NREREJ}(1)<0) and uses
          pseudopotential (\ref{eq:1})
\item four coupling constants $W^{(n)}$, $B^{(n)}$, $H^{(n)}$,
          and $M^{(n)}$ per line (for \tv{NREREJ}(1)>0), and uses local
          pseudopotential (\ref{eq:20}).
\end{itemize}
          $N$=2*|\tv{NREREJ}(1)| denotes
          the order of expansion (the maximum value of $n$) and
          $a$=\tv{REGWID} denotes the width of the Gaussian
          formfactor $g_a(\bm{r})$, see
          Section~\ref{subsec:Regularized}.

          For \tv{NREREJ}(1)<0, the codes reads the coupling
          constants corresponding to terms defined in Eqs.~(42)--(54)
          of Ref.~\cite{(Rai14a)}, with the exception of coupling
          constants corresponding to terms that depend on the
          $\hat{T}_3$ operator, see Eq.~(41) of Ref.~\cite{(Rai14a)}.
          The latter terms were not yet implemented. Consecutive
          lines of input are numbered by index $k=1,\ldots{i}$ and
          correspond to the coupling constants defined by indices $n$
          and $j$ as:

          \begin{equation}\begin{array}{rrr|rrr|rrr|rrr}
          \hline
            k  &  n  &  j &  k  &  n  &  j &  k  &  n  &  j &  k  &  n  &   j \\
          \hline
            1  &  0  &  1 &  2  &  2  &  1 &  4  &  4  &  1 &   7 &  6  &   1 \\
               &     &    &  3  &  2  &  2 &  5  &  4  &  2 &   8 &  6  &   2 \\
               &     &    &     &     &    &  6  &  4  &  3 &   9 &  6  &   3 \\
               &     &    &     &     &    &     &     &    &  10 &  6  &   4 \\
          \hline
          \end{array}\end{equation}

          The number of lines read $i$ must be consistent with the
          order of expansion, that is, for \tv{NREREJ}(1)<0 or
          \tv{NREREJ}(1)$\geq$0 there must be exactly
          $i=(N+2)(N+4)/8$ or $i=(N+2)/2$ lines read, respectively.

          For \tv{IREREJ}(1)=0, no lines with parameters are read,
          and the code uses the local higher-order pseudopotential
          (\ref{eq:20}) with coupling constants $W^{(n)}$, $B^{(n)}$,
          $H^{(n)}$, and $M^{(n)}$ derived from the Gogny interaction
          up to order $N$. The methodology and equations developed in
          Ref.~\cite{(Dob12b)} are then used. For \tv{IREREJ}(1)=0,
          \tv{NREREJ}(1)<0 is not allowed.

          In a given run of the code \pr{hfodd}, \begin{itemize}\item keyword
          \tk{REGUCOUPLI} must not be simultaneously used with
          keyword \tk{REGUL\_TXYZ},\item all read values of array
          \tv{IREREJ} must be the same and are internally used as
          variable \tv{IREREG},\item all read values of array
          \tv{NREREJ} must be the same and are internally used as
          variable \tv{N3LORD}.\end{itemize} Internally, the code uses array
          \tv{REGVCC}=\tv{REJVCC}. To inform the code on whether the
          coupling constants had been read from the input data file,
          all elements of array \tv{IREREJ} are predefined to 100.

          In version {(v\codeversion)} of the code \pr{hfodd},
          calculations with the nonlocal pseudopotential
          (\ref{eq:1}) or local pseudopotential (\ref{eq:20}) are
          implemented up to order $N=4$ (N2LO) or $N=6$ (N3LO),
          respectively. Consequently, the only allowed values are
          $-2\leq\tv{N3LORD}\leq3$.

\key{REGCOUPERR} \hspace*{-20mm}\begin{tabular}[t]{lcl}
                   100,  0,           &=& \tv{IREREJ}(2),\tv{NREREJ}(2)     \\
                    0.,  0.,  0.,  0. &=& \tv{REJERR}(1,1),\tv{REJERR}(1,2),\tv{REJERR}(1,3),\tv{REJERR}(1,4) \\
                    0.,  0.,  0.,  0. &=& \tv{REJERR}(2,1),\tv{REJERR}(2,2),\tv{REJERR}(2,3),\tv{REJERR}(2,4) \\
                    \multicolumn{3}{c}{\dotfill} \\
                    0.,  0.,  0.,  0. &=& \tv{REJERR}(i,1),\tv{REJERR}(i,2),\tv{REJERR}(i,3),\tv{REJERR}(i,4) \\
                   \end{tabular}

\keyspace

\key{REGCOUPFAC}   \hspace*{-20mm}\begin{tabular}[t]{lcl}
                   100,  0,           &=& \tv{IREREJ}(3),\tv{NREREJ}(3)     \\
                    0.,  0.,  0.,  0. &=& \tv{REJFAC}(1,1),\tv{REJFAC}(1,2),\tv{REJFAC}(1,3),\tv{REJFAC}(1,4) \\
                    0.,  0.,  0.,  0. &=& \tv{REJFAC}(2,1),\tv{REJFAC}(2,2),\tv{REJFAC}(2,3),\tv{REJFAC}(2,4) \\
                    \multicolumn{3}{c}{\dotfill} \\
                    0.,  0.,  0.,  0. &=& \tv{REJFAC}(i,1),\tv{REJFAC}(i,2),\tv{REJFAC}(i,3),\tv{REJFAC}(i,4) \\
                   \end{tabular}

\keyspace

\key{REGUL\_TXYZ} \hspace*{-20mm}\begin{tabular}[t]{lcl}
                   100,  0,  1.       &=& \tv{IREREJ}(4),\tv{NREREJ}(4),\tv{REGWID}     \\
                    0.,  0.,  0.,  0. &=& \tv{REJTCC}(1,1),\tv{REJTCC}(1,2),\tv{REJTCC}(1,3),\tv{REJTCC}(1,4) \\
                    0.,  0.,  0.,  0. &=& \tv{REJTCC}(2,1),\tv{REJTCC}(2,2),\tv{REJTCC}(2,3),\tv{REJTCC}(2,4) \\
                    \multicolumn{3}{c}{\dotfill} \\
                    0.,  0.,  0.,  0. &=& \tv{REJTCC}(i,1),\tv{REJTCC}(i,2),\tv{REJTCC}(i,3),\tv{REJTCC}(i,4) \\
                   \end{tabular}

\keyspace

\noindent The three keywords above allow for reading the coupling constants
          of pseudopotentials (\ref{eq:1}) or (\ref{eq:20}) in the format
          analogous to that used for keyword \tk{REGUCOUPLI}. Array \tv{REJTCC}
          contains the strength parameters of Eq.~(\ref{eq:2}), that is,
          \begin{equation}\begin{array}{rcl}
          \tv{REGVCC}(:,1)&=&\tv{REJTCC}(:,1) \\
          \tv{REGVCC}(:,2)&=&\tv{REJTCC}(:,1)*\tv{REJTCC}(:,2) \\
          \tv{REGVCC}(:,3)&=&\tv{REJTCC}(:,1)*\tv{REJTCC}(:,3) \\
          \tv{REGVCC}(:,4)&=&\tv{REJTCC}(:,1)*\tv{REJTCC}(:,4) \\
          \end{array}\end{equation}
          Arrays \tv{REJERR} and \tv{REJFAC} modify array \tv{REGVCC} as
          \begin{equation}\begin{array}{rcl}
          \tv{REGVCC}(:,:)&=&\tv{REGVCC}(:,:)+\tv{REJERR}(:,:)*\tv{REJFAC}(:,:) \\
          \end{array}\end{equation}
          Such a modification of array \tv{REGVCC} is performed
          regardless of whether keyword \tk{REGUL\_TXYZ} or
          \tk{REGUCOUPLI} was used to define it. In a given run of
          the code \pr{hfodd}, \begin{itemize}\item keyword
          \tk{REGUCOUPLI} must not be simultaneously used with
          keyword \tk{REGUL\_TXYZ},\item all read values of array
          \tv{IREREJ} must be the same and are internally used as
          variable \tv{IREREG},\item all read values of array
          \tv{NREREJ} must be the same and are internally used as
          variable \tv{N3LORD}.\end{itemize} To inform the code on
          whether the coupling constants had been read from the
          input data file, all elements of array \tv{IREREJ} are
          predefined to 100.

\subsubsection{Separable terms. \\[2ex]\hspace*{-1em}}
\label{subsubsec:Separable}

\key{SEPARGAUSS}  0 = \tv{I\_SEPA}

\keyspace

\noindent For \tv{I\_SEPA}>0, the average mean-field energies of the
          separable pseudopotential (\ref{eq:101}) are calculated.
          For \tv{I\_SEPA}=2 or 3, the corresponding direct mean
          fields are included in the self-consistent mean field. For
          \tv{I\_SEPA}=2 or 4, the corresponding exchange mean fields
          are included in the self-consistent mean field. Altogether,
          \tv{I\_SEPA}=1 demands calculations of contributions to
          energy only, whereas \tv{I\_SEPA}=2 demands full
          self-consistent calculations with both direct and exchange
          mean fields included. For \tv{I\_SEPA}=0, the mean field
          corresponding to the separable pseudopotential is ignored.

          For \tv{I\_SEPA}>0 and \tv{IPAHFB}>1, the code issues a
          warning to the effect that, unless the zero-range pairing
          strengths are explicitly set to zero, see
          Section~IV-3.1~\cite{(Dob04d)}, the corresponding pairing
          still will be active. For \tv{NOZEPA}=1, see
          Section~\ref{subsubsec:Miscellaneous}, the zero-range
          pairing is neglected, and the warning is not printed.
          In version {(v\codeversion)} of the code
          \pr{hfodd}, \tv{I\_SEPA}>0 still requires \tv{IPNMIX}=0,
          \tv{IRENMA}=0, and \tv{IBROYD}=0.

\key{SEPAR\_PAIR}  0 = \tv{ISEPPA}

\keyspace

\noindent For \tv{ISEPPA}>0, the average pairing energies of the
          separable pseudopotential (\ref{eq:101}) are calculated.
          For \tv{ISEPPA}=2, the corresponding pairing fields are
          included in the self-consistent pairing field. Altogether,
          \tv{ISEPPA}=1 demands calculations of contributions to the
          pairing energy, whereas \tv{I\_SEPA}=2 demands full
          self-consistent calculations with pairing fields included.
          For \tv{I\_SEPA}=0, the pairing contribution of the
          separable pseudopotential is ignored. \tv{ISEPPA}>0
          requires \tv{IPAHFB}>0. In version {(v\codeversion)} of the code
          \pr{hfodd}, \tv{ISEPPA}>0 still requires \tv{IPNMIX}=0,
          \tv{IRENMA}=0, and \tv{IBROYD}=0.

\key{SEPCOUPLI} \hspace*{-20mm}\begin{tabular}[t]{lcl}
                    0,  0             &=& \tv{IVISEP},\tv{N3SERD}     \\
                    0.,  0.,  0.,  0. &=& \tv{SEPVIC}(1,1),\tv{SEPVIC}(1,2),\tv{SEPVIC}(1,3),\tv{SEPVIC}(1,4) \\
                   \end{tabular}
\keyspace

\noindent After reading the first line, the code reads one line with
          the four coupling constants $\tilde{W}$, $\tilde{B}$,
          $\tilde{H}$, and $\tilde{M}$ and uses separable
          pseudopotential (\ref{eq:101}). In version
          {(v\codeversion)} of the code \pr{hfodd}, only values of
          $i$=\tv{IVISEP}=0 or 1 and $n$=\tv{N3SERD}=0 are allowed;
          other values may become available after higher-order
          derivative terms are implemented, in analogy to those of
          the regularised pseudopotential (\ref{eq:1}). Unless
          keyword \tk{SEPAR\_FORM} with \tv{NUSEGA}>0 is used, the
          use of keyword \tk{SEPCOUPLI} implies that the formfactor
          (\ref{eq:102}) is composed of one Gaussian only, with the
          default values of $K$=\tv{NUSEGA}=1,
          $A_1$=\tv{SEPGAU}(1)=1, and $a_1$=\tv{SEPWID}(1)=1.
          Internally, the code uses variable \tv{IRESEP}=\tv{IVISEP}
          and array \tv{SEPVCC}=\tv{SEPVIC}. In a given run of the
          code \pr{hfodd}, keyword \tk{SEPCOUPLI} must not be
          simultaneously used with keyword \tk{SEP\_TXYZ}.

\key{SEP\_TXYZ} \hspace*{-20mm}\begin{tabular}[t]{lcl}
                    0,  0             &=& \tv{IVISEP},\tv{N3SERD}     \\
                    0.,  0.,  0.,  0. &=& \tv{SEPTIC}(1,1),\tv{SEPTIC}(1,2),\tv{SEPTIC}(1,3),\tv{SEPTIC}(1,4) \\
                   \end{tabular}

\keyspace

\key{SEPCOUPERR} \hspace*{-20mm}\begin{tabular}[t]{lcl}
                    0                 &=& \tv{IERSEP}     \\
                    0.,  0.,  0.,  0. &=& \tv{SEPERR}(1,1),\tv{SEPERR}(1,2),\tv{SEPERR}(1,3),\tv{SEPERR}(1,4) \\
                   \end{tabular}

\keyspace

\key{SEPCOUPFAC} \hspace*{-20mm}\begin{tabular}[t]{lcl}
                    0                 &=& \tv{IFASEP}     \\
                    0.,  0.,  0.,  0. &=& \tv{SEPFAC}(1,1),\tv{SEPFAC}(1,2),\tv{SEPFAC}(1,3),\tv{SEPFAC}(1,4) \\
                   \end{tabular}

\keyspace

\noindent The three keywords above allow for reading the coupling constants
          of the separable pseudopotential (\ref{eq:101}) in the format
          analogous to that used for keyword \tk{SEPCOUPLI}. Array \tv{SEPTIC}
          contains strength parameters $\tilde{t}$, $\tilde{x}$,
          $\tilde{y}$, and $\tilde{z}$ defining the coupling constants as
          $\tilde{W}=\tilde{t}$,
          $\tilde{B}=\tilde{t}\tilde{x}$,
          $\tilde{H}=\tilde{t}\tilde{y}$, and
          $\tilde{M}=\tilde{t}\tilde{z}$, that is,
          \begin{equation}\begin{array}{rcl}
          \tv{SEPVIC}(:,1)&=&\tv{SEPTIC}(:,1) \\
          \tv{SEPVIC}(:,2)&=&\tv{SEPTIC}(:,1)*\tv{SEPTIC}(:,2) \\
          \tv{SEPVIC}(:,3)&=&\tv{SEPTIC}(:,1)*\tv{SEPTIC}(:,3) \\
          \tv{SEPVIC}(:,4)&=&\tv{SEPTIC}(:,1)*\tv{SEPTIC}(:,4) \\
          \end{array}\end{equation}
          Internally, the code uses variable
          \tv{IRESEP}=\tv{IVISEP} and array \tv{SEPVCC}=\tv{SEPVIC}.
          Arrays \tv{SEPERR} and \tv{SEPFAC} modify array \tv{SEPVCC} as
          \begin{equation}\begin{array}{rcl}
          \tv{SEPVCC}(:,:)&=&\tv{SEPVCC}(:,:)+\tv{SEPERR}(:,:)*\tv{SEPFAC}(:,:) \\
          \end{array}\end{equation}
          Such a modification of array \tv{SEPVCC} is performed
          regardless of whether keyword \tk{SEP\_TXYZ} or
          \tk{SEPCOUPLI} was used to define it. In version
          {(v\codeversion)} of the code \pr{hfodd}, only values of
          $i$=\tv{IERSEP}=\tv{IFASEP}=\tv{IVISEP}=0 or 1 and
          $n$=\tv{N3SERD}=0 are allowed. In a given run of the code
          \pr{hfodd}, keyword \tk{SEP\_TXYZ} must not be
          simultaneously used with keyword \tk{SEPCOUPLI}.

\key{SEPAR\_FORM} \hspace*{-20mm}\begin{tabular}[t]{lcl}
                    1                     &=& \tv{NUSEGA}     \\
                    1.,  1.,  \ldots,  1. &=& \tv{SEPWID}(1),\tv{SEPWID}(2),\ldots,\tv{SEPWID}(\tv{NUSEGA}) \\
                    1.,  0.,  \ldots,  0. &=& \tv{SEPGAU}(1),\tv{SEPGAU}(2),\ldots,\tv{SEPGAU}(\tv{NUSEGA}) \\
                   \end{tabular}

\keyspace

\noindent After reading the first line, for $K$=\tv{NUSEGA}>0 the code reads two lines
          with widths $a_k$=\tv{SEPWID}(k) and amplitudes $A_k$=\tv{SEPGAU}(k)
          of Gaussians that define formfactor (\ref{eq:102}) of the separable
          pseudopotential (\ref{eq:101}).

\subsection{Symmetries}
\label{subsubsec:Symmetries}

\key{HFB2HF} 0,   0,   =   \tv{IPA2HF}(0),\tv{IPA2HF}(1)

\keyspace

\key{GAP2HF} 0.,  0.,  =   \tv{DEL2HF}(0),\tv{DEL2HF}(1)

\keyspace

\noindent The two keywords above allow for the use of a hybrid
          method of calculations, where the HF method is used for
          neutrons (protons), \tv{IPA2HF}(0)=1(0), and the HFB
          method is  used for protons (neutrons),
          \tv{IPA2HF}(0)=0(1). For \tv{IPA2HF}(0)=\tv{IPA2HF}(1)=0,
          the hybrid method is inactive and the code proceeds as
          dictated by other keywords handling the HF/HFB method,
          whereas for \tv{IPA2HF}(0)=\tv{IPA2HF}(1)=1, the HF method
          is enforced for both neutrons and protons, irrespective of
          what is dictated by other keywords handling the HF/HFB
          method.

          For \tv{IPA2HF}(0)=2 or \tv{IPA2HF}(1)=2, the HFB
          calculations requested for neutrons or protons will {\em
          during the iterations} automatically switch over to HF as
          soon as the neutron or proton pairing gap goes below
          $\tv{DEL2HF}(0)$ or $\tv{DEL2HF}(1)$, respectively. The
          user is responsible for properly setting the keywords
          handling the HF method before the HFB run is started,
          because the correctness and consistency of these keywords
          would not be pre-tested. \tv{IPA2HF}(0)>0 or \tv{IPA2HF}(1)>0
          requires \tv{IPAHFB}(0)>0.

\subsection{Symmetry restoration}
\label{subsubsec:Restoration}

\key{PROJPARNUM} 0, 1 = \tv{IPRNUM}, \tv{NPNKNO}

\keyspace

\noindent For \tv{IPRNUM}=1(2) and \tv{NPNKNO}>1, and for diagonal
          (non-diagonal) GCM kernels, see
          Section~VI-3.2~\cite{(Dob09g)}, the code performs
          projection on total particle number
          $A$=\tv{IN\_FIX}+\tv{IZ\_FIX}. The number of Gauss-Tchebyschev
          points used to perform the integration over gauge angle
          $\phi$ covering the domain of $0\leq\phi<\pi$ is defined by
          \tv{NPNKNO}. \tv{IPRNUM}>0 requires that it is equal to all
          other nonzero projection switches: \tv{IPRROT},
          \tv{IPRISO}, \tv{IPRVEC}, and \tv{IPRPTY}, see
          Section~\ref{subsubsec:ISOSTZ}.

\key{PROJVECNUM} 0, 1 = \tv{IPRVEC}, \tv{NTZKNO}

\keyspace

\noindent For \tv{IPRVEC}=1(2) and \tv{NTZKNO}>1, and for diagonal
          (non-diagonal) GCM kernels, see
          Section~VI-3.2~\cite{(Dob09g)}, the code performs
          projection on doubled $z$-component of the isospin.
          2$T_z$=\tv{IN\_FIX}-\tv{IZ\_FIX}. The number of Gauss-Tchebyschev
          points used to perform the integration over gauge angle
          $\phi_T$ covering the domain of $0\leq\phi_T<\pi$ is defined by
          \tv{NTZKNO}. \tv{IPRNUM}>0 requires that it is equal to all
          other nonzero projection switches: \tv{IPRROT},
          \tv{IPRISO}, \tv{IPRNUM}, and \tv{IPRPTY}, see
          Section~\ref{subsubsec:ISOSTZ}.

\key{PROJPARITY} 0, 0, +1 = \tv{IPRPTY}, \tv{NPAKNO}, \tv{IPAPRO}

\keyspace

\noindent For \tv{IPRPTY}=1(2) and \tv{NPAKNO}=2, and for diagonal
          (non-diagonal) GCM kernels, see
          Section~VI-3.2~\cite{(Dob09g)}, the code performs parity
          projection onto the positive-parity (for \tv{IPAPRO}=+1) or
          negative-parity (for \tv{IPAPRO}=$-$1) states.
          \tv{IPRPTY}>0 requires that it is equal to all other
          nonzero projection switches: \tv{IPRROT}, \tv{IPRISO},
          \tv{IPRNUM}, and \tv{IPRVEC}, see
          Section~\ref{subsubsec:ISOSTZ}.

\key{ONISHI} 0 = \tv{IONISH}

\keyspace

\noindent For \tv{IONISH}=0 or 1, the code uses the Pfaffian or
          Onishi formula, respectively, see
          Section~\ref{subsubsec:Pfaffian}, to compute the overlap
          kernels between the HFB wave functions involved in the
          symmetry projection (keywords \tk{PROJECTGCM},
          \tk{PROJPARITY}, \tk{PROJPARNUM}, or \tk{PROJVECNUM}). \tv{IPRGCM}>1 with
          \tv{IPAHFB}>1 and \tv{IONISH}=0 requires \tv{ITWOBA}=1 and
          \tv{NUQEVE}=1. In version {(v\codeversion)} of the code
          \pr{hfodd}, \tv{IPRGCM}>1 with \tv{IPAHFB}>1 and
          \tv{IONISH}=0 still requires \tv{ISIMPY}=0 and
          \tv{ISIQTY}=0.

\key{PROJ\_J2\_T2} 0,0 = \tv{KETAJ2}, \tv{KETAT2}

\keyspace

\noindent For \tv{KETAJ2}=1 (\tv{KETAT2}=1) and \tv{IPRGCM}>0, the code computes the
          expectation value of the square of total angular momentum
          (total isospin) in the projected states. Such a calculation
          is performed to control the precision of the
          angular-momentum (isospin) projection. \tv{KETAJ2}=1
          (\tv{KETAT2}=1) is ignored unless  \tv{NUBKNO}=1
          (\tv{NBTKNO}=1). Either \tv{KETAJ2}=1 or \tv{KETAT2}=1
          requires \tv{ISAKER}=0 or 2.

\key{KERNINVERS} 0,  0, =  \tv{IKEINV},\tv{IKEKAR}

\keyspace

\noindent For \tv{IKEINV}>0 or \tv{IKEKAR}>0, and for \tv{IPRGCM}=1,
          the "right" wave function of the kernel calculated within the
          diagonal GCM mode, see Section~VI-3.2~\cite{(Dob09g)}, is
          initially transformed according to one of the
          $D^{\text{T}}_{\text{2h}}$ symmetry operations enumerated
          for variables \tv{INIINV} and \tv{INIKAR} under keyword
          \tk{INI\_INVERS} in Section~VI-3.2~\cite{(Dob09g)}.
          Compared to the operations requested by keyword
          \tk{INI\_INVERS}, which generate the transformed
          wavefunction that can later be used by employing keyword
          \tk{PROJECTGCM} in the non-diagonal GCM mode, see
          Section~VI-3.2~\cite{(Dob09g)}, keyword \tk{KERNINVERS}
          ensures that the phase relations between the "right" and
          "left" wavefunctions are properly maintained.

\key{NOBLOLIPKI} 0,  0, =  \tv{LIPNON},\tv{LIPNOP}

\keyspace

\noindent For \tv{LIPNON}=1 or \tv{LIPNOP}=1, the blocked neutron or
          proton orbitals are excluded from the calculation of the
          neutron or proton Lipkin parameters $\lambda_2$,
          respectively. Since the occupation probabilities of the
          blocked states are by definition equal to 1, their
          contributions to particle-number fluctuations should not be,
          in principle, counted. \tv{LIPNON}=1 requires \tv{LIPKIN}=1 and
          \tv{LIPNOP}=1 requires \tv{LIPKIP}=1.

\key{PROJE\_DENS} 0 =  \tv{IDENSU}

\keyspace

\noindent For \tv{IDENSU}=1, the particle-number projection is
performed with ignored gauge-angle dependence of the density in the
p-h and p-p density-dependent terms.

\subsection{Configurations}
\label{subsubsec:Configurations}

\key{AXIALIZE}  0 =  \tv{IAXIAP}

\keyspace

\noindent For |\tv{IAXIAP}|=1, the code axializes, see
          Section~\ref{subsec:Axialization}, the particle-hole and
          pairing (for \tv{IPAHFB}>0) mean-fields. For
          \tv{IAXIAP}=$-$1, the code in addition axializes the
          particle-hole density matrix and pairing tensor (for
          \tv{IPAHFB}>0). For |\tv{IAXIAP}|=1, \tv{ICONTI}=1, and
          \tv{IFCONT}=0, the code issues a warning to the effect that
          a smooth continuation of axialized wavefunctions may
          require continuation from fields, that is, \tv{IFCONT}=1,
          see Section~VI-3.8~\cite{(Dob09g)}. In version
          {(v\codeversion)} of the code \pr{hfodd}, |\tv{IAXIAP}|=1
          still requires \tv{IGOGPA}=0 and \tv{IAXIAP}=$-$1 is incompatible
          with the angular-momentum projection (\tv{NUBKNO}>1).

\key{VACNONANEU}    \begin{tabular}[t]{lcl}
                    0, 1         &=& \tv{NLSIZN},\tv{MXALIN}           \\
                    0, \ldots, 0 &=& \tv{LALSIZ}($-$\tv{MXALIN},0),    \\
                                  && \tv{LALSIZ}($-$\tv{MXALIN}+2,0),  \\
                                  &&\multicolumn{1}{c}{\dotfill}       \\
                                  && \tv{LALSIZ}(+\tv{MXALIN}$-$2,0),  \\
                                  && \tv{LALSIZ}(+\tv{MXALIN},0)       \\
                   \end{tabular}

\keyspace

\noindent After reading the first line, the code reads the second line that contains \tv{MXALIN}+1
          numbers \tv{LALSIZ}(:,0) of neutrons in the $\Omega$-blocks, from $\Omega$=$-$\tv{MXALIN}/2
          to $\Omega$=+\tv{MXALIN}/2, see Section~\ref{subsec:partitions}.
          |\tv{NLSIZN}|=1, 2, or 3 stands for the
          Cartesian direction of $x$, $y$, or $z$, respectively. For \tv{NLSIZN}>0,
          the code distributes neutrons in the $\Omega$-blocks according to the
          values of \tv{LALSIZ}(i,0).
          For \tv{LALSIZ}(i,0)>0, the |\tv{LALSIZ}(i,0)| lowest neutron
          states are occupied in the block $\Omega$=$i/2$.
          For \tv{LALSIZ}(i,0)<0, the |\tv{LALSIZ}(i,0)|$-$1 lowest neutron
          states are occupied in the block $\Omega$=$i/2$, the state number |\tv{LALSIZ}(i,0)|
          is kept empty, and the state number |\tv{LALSIZ}(i,0)|+1 is kept occupied.
          For \tv{NLSIZN}<0, the code does not fix occupations in the $\Omega$-blocks
          but determines and prints the distribution of neutrons across the $\Omega$-blocks.
          \tv{MXALIN} must be odd. For
          \tv{IPAIRI}=0, \tv{IPNMIX}=0,
          \tv{ISIMPY}=0, \tv{ISIQTY}=0,
          \tv{IVACUM}=0, and \tv{NLSIZN}>0, the sum of numbers of neutrons
          in all $\Omega$-blocks, that is, the sum of |\tv{LALSIZ}(:,0)|, must be equal
          to the number of neutrons given by \tv{IN\_FIX}.
          |\tv{NLSIZN}|>0 requires \tv{ISIMPY}=0 and \tv{ISIQTY}=0.

\key{VACNONAPRO}    \begin{tabular}[t]{lcl}
                    0, 1         &=& \tv{NLSIZP},\tv{MXALIP}           \\
                    0, \ldots, 0 &=& \tv{LALSIZ}($-$\tv{MXALIP},1),    \\
                                  && \tv{LALSIZ}($-$\tv{MXALIP}+2,1),  \\
                                  &&\multicolumn{1}{c}{\dotfill}       \\
                                  && \tv{LALSIZ}(+\tv{MXALIP}$-$2,1),  \\
                                  && \tv{LALSIZ}(+\tv{MXALIP},1)       \\
                   \end{tabular}

\keyspace

\noindent The same as in keyword \tk{VACNONANEU} but for protons.

\key{VACPARANEU}    \begin{tabular}[t]{lcl}
                    0, 1         &=& \tv{NLSIQN},\tv{MXALIN}             \\
                    0, \ldots, 0 &=& \tv{LALSIQ}($-$\tv{MXALIN},0,0),    \\
                                  && \tv{LALSIQ}($-$\tv{MXALIN}+2,0,0),  \\
                                  &&\multicolumn{1}{c}{\dotfill}         \\
                                  && \tv{LALSIQ}(+\tv{MXALIN}$-$2,0,0),  \\
                                  && \tv{LALSIQ}(+\tv{MXALIN},0,0)       \\
                    0, \ldots, 0 &=& \tv{LALSIQ}($-$\tv{MXALIN},1,0),    \\
                                  && \tv{LALSIQ}($-$\tv{MXALIN}+2,1,0),  \\
                                  &&\multicolumn{1}{c}{\dotfill}         \\
                                  && \tv{LALSIQ}(+\tv{MXALIN}$-$2,1,0),  \\
                                  && \tv{LALSIQ}(+\tv{MXALIN},1,0)       \\
                   \end{tabular}

\keyspace

\noindent After reading the first line, the code reads two lines that each contain \tv{MXALIN}+1
          numbers of neutrons \tv{LALSIQ}(:,0,0) and \tv{LALSIQ}(:,1,0) in
          the positive-parity and negative-parity
          $\Omega$-blocks, respectively, from $\Omega$=$-$\tv{MXALIN}/2
          to $\Omega$=+\tv{MXALIN}/2, see Section~\ref{subsec:partitions}.
          |\tv{NLSIQN}|=1, 2, or 3 stands for the
          Cartesian direction of $x$, $y$, or $z$, respectively. For \tv{NLSIQN}>0,
          the code distributes neutrons in the positive-parity and negative-parity
          $\Omega$-blocks according to the
          values of \tv{LALSIQ}(i,j,0), where j=0(1) stands for the positive (negative) parity.
          For \tv{LALSIQ}(i,j,0)>0, the |\tv{LALSIQ}(i,j,0)| lowest neutron
          states are occupied in the block $\Omega$=$i/2$ for a given parity.
          For \tv{LALSIQ}(i,j,0)<0, the |\tv{LALSIQ}(i,j,0)|$-$1 lowest neutron
          states are occupied in the block $\Omega$=$i/2$ for a given parity,
          the state number |\tv{LALSIQ}(i,j,0)|
          is kept empty, and the state number |\tv{LALSIQ}(i,j,0)|+1 is kept occupied.
          For \tv{NLSIQN}<0, the code does not fix occupations in the $\Omega$-blocks
          but determines and prints the distribution of neutrons across the $\Omega$-blocks
          of both parities.
          \tv{MXALIN} must be odd. For
          \tv{IPAIRI}=0, \tv{IPNMIX}=0,
          \tv{ISIMPY}=0, \tv{ISIQTY}=1,
          \tv{IVACUM}=0, and \tv{NLSIQN}>0, the sum of numbers of
          neutrons in all $\Omega$-blocks of a given parity, that is,
          the sum of |\tv{LALSIQ}(:,j,0)|, must be equal to the
          number of neutrons given by \tv{KVASIQ}(j,0),
          see~Section~IV-3.3~\cite{(Dob04d)}.\footnote{In Ref.~\cite{(Dob04d)}~p.~175,
          description of keywords \tk{VACPAR\_NEU} and \tk{VACPAR\_PRO}
          should refer to variables \tv{KVASIQ}(0,0),\tv{KVASIQ}(1,0)
          and \tv{KVASIQ}(0,1),\tv{KVASIQ}(1,1), respectively.} |\tv{NLSIQN}|>0
          requires \tv{ISIMPY}=0 and \tv{ISIQTY}=1.

\key{VACPARAPRO}     \begin{tabular}[t]{lcl}
                    0, 1         &=& \tv{NLSIQP},\tv{MXALIN}             \\
                    0, \ldots, 0 &=& \tv{LALSIQ}($-$\tv{MXALIN},0,1),    \\
                                  && \tv{LALSIQ}($-$\tv{MXALIN}+2,0,1),  \\
                                  &&\multicolumn{1}{c}{\dotfill}         \\
                                  && \tv{LALSIQ}(+\tv{MXALIN}$-$2,0,1),  \\
                                  && \tv{LALSIQ}(+\tv{MXALIN},0,1)       \\
                    0, \ldots, 0 &=& \tv{LALSIQ}($-$\tv{MXALIN},1,1),    \\
                                  && \tv{LALSIQ}($-$\tv{MXALIN}+2,1,1),  \\
                                  &&\multicolumn{1}{c}{\dotfill}         \\
                                  && \tv{LALSIQ}(+\tv{MXALIN}$-$2,1,1),  \\
                                  && \tv{LALSIQ}(+\tv{MXALIN},1,1)       \\
                   \end{tabular}

\keyspace

\noindent   The same as in keyword \tk{VACPARANEU} but for protons.

\key{VACLASTORB} 0, 0, =  \tv{LASTAN},\tv{LASTAP}

\keyspace

\noindent For \tv{LASTAN}=1 or \tv{LASTAP}=1, calculations are
          performed with only one, highest-energy neutron or proton
          orbital, respectively, occupied in each $\Omega$-block, see
          Section~\ref{subsec:partitions}. This option can be used to
          perform the angular-momentum projection of a
          single-orbital, after the code is restarted from a
          converged solution. For \tv{LASTAN}=1, the option od using
          negative values of \tv{LALSIZ} or \tv{LALSIQ}, see keywords
          \tk{VACNONANEU} or \tk{VACPARANEU}, respectively, is not
          allowed. The same rule applies for protons.

\key{FILNON\_NEU} 2, 1, 0  = \tv{KPFILZ}(0),\tv{KHFILZ}(0),\tv{KOFILZ}(0)

\keyspace

\noindent Keyword \tk{FILNON\_NEU} is an analogue of keyword
          \tk{FILSIG\_NEU}, see Section~VI-3.2~\cite{(Dob09g)}, and
          demands calculations performed within the filling
          approximation applied to neutrons in the no-symmetry case. Variables
          \tv{KPFILZ}(0) and \tv{KHFILZ}(0) contain indices of
          particle (empty) and hole (occupied) states, respectively.
          Variable \tv{KOFILZ}(0) contains the number of particles
          put into the states between \tv{KHFILZ}(0) and
          \tv{KPFILZ}(0) by using for them partial occupation factors
          of \tv{KOFILZ}(0)/(\tv{KPFILZ}(0)$-$\tv{KHFILZ}(0)+1). For
          \tv{KOFILZ}(0) = 0, the filling approximation is inactive.
          \tv{KOFILZ}(0)>0 is incompatible with \tv{IPAIRI}=1,
          \tv{IFLIPI}$\neq$0, or \tv{KOFLIZ}(0)$\neq$0.

\key{FILNON\_PRO} 2, 1, 0  = \tv{KPFILZ}(1),\tv{KHFILZ}(1),\tv{KOFILZ}(1)

\keyspace

\noindent Same as for keyword \tk{FILNON\_NEU} but it
          demands calculations performed within the filling
          approximation applied to protons in the no-symmetry case.

\key{MBLOCSIZ\_N}  \begin{tabular}[t]{lcl}
                    0     &=& \tv{NBBLOC}             \\
                    1,  0 &=& \tv{INSIZN}(1),\tv{IDSIZN}(1),      \\
                    1,  0 &=& \tv{INSIZN}(2),\tv{IDSIZN}(2),      \\
                          &&\multicolumn{1}{c}{\dotfill}         \\
                    1,  0 &=& \tv{INSIZN}(\tv{NBBLOC}),\tv{IDSIZN}(\tv{NBBLOC})    \\
                   \end{tabular}

\keyspace

\noindent Keyword \tk{MBLOCSIZ\_N} generalizes the quasiparticle
          blocking requested by keyword \tk{BLOCKSIZ\_N} in the case
          of no symmetries, see~Section~IV-3.3~\cite{(Dob04d)} to
          multi-quasiparticle blocking, see
          Section~\ref{subsubsec:Multi-particle}. After reading the
          first line, the code reads \tv{NBBLOC} pairs of data,
          \tv{INSIZN}($i$) and \tv{IDSIZN}($i$), for
          $i$=1,\ldots,\tv{NBBLOC}. For \tv{IDSIZN}($i$)=+1 or $-1$,
          the blocked quasiparticle state is selected by having the
          largest overlap with the \tv{INSIZN}($i$)th neutron
          single-particle eigenstate of the HFB mean-field Routhian
          or with its time-reversed partner, respectively. Note that
          for rotating states, the time-reversed eigenstate is not
          necessarily an eigenstate of the Routhian. For
          \tv{IDSIZN}($i$)=0, the blocking of the $i$th
          quasiparticle is omitted. For any $i$, |\tv{IDSIZN}($i$)|=1
          requires \tv{ISIMPY}=0, \tv{IPARTY}=0, \tv{IPAHFB}=1, and
          \tv{IROTAT}=1. In a given run of the code \pr{hfodd},
          keyword \tk{MBLOCSIZ\_N} must not be simultaneously used
          with keyword \tk{BLOCKSIZ\_N}.

\key{MBLOCSIZ\_P}  \begin{tabular}[t]{lcl}
                    0     &=& \tv{NBBLOC}             \\
                    1,  0 &=& \tv{INSIZP}(1),\tv{IDSIZP}(1),      \\
                    1,  0 &=& \tv{INSIZP}(2),\tv{IDSIZP}(2),      \\
                          &&\multicolumn{1}{c}{\dotfill}         \\
                    1,  0 &=& \tv{INSIZP}(\tv{NBBLOC}),\tv{IDSIZP}(\tv{NBBLOC})    \\
                   \end{tabular}

\keyspace

\noindent Same as for keyword \tk{MBLOCSIZ\_N} but for the proton multi-quasiparticle blocking.

\key{NUMBCUTOFF}  0, 0  = \tv{NCUTOF}(0),\tv{NCUTOF}(1)

\keyspace

\noindent For \tv{NCUTOF}(0)>0 and/or \tv{NCUTOF}(1)>0, the
          numbers of lowest neutron and/or proton single-particle
          states used in the two-basis method, see
          Section~VII-2.2.1~\cite{(Sch12c)}, are limited to
          \tv{NCUTOF}(0) and/or \tv{NCUTOF}(1), respectively. This
          option overrides the energy cutoff specified by variable
          \tv{ECUTOF} read under keyword \tk{CUTOFF},
          see~Section~IV-3.1~\cite{(Dob04d)}. This option allows for
          the calculation of transition densities when it is performed
          for different left and right HFB states, that is for
          \tv{IPRGCM}>1. Indeed, such type of calculation requires
          that the numbers of quasiparticle states defining the left
          and right HFB states are equal. \tv{NCUTOF}(0)>0 and/or
          \tv{NCUTOF}(1)>0 is incompatible with \tv{ITWOBA}=0 (for now),
          \tv{LIMQUA}=1, or \tv{LAMCUT}=1.

\subsection{Numerical parameters}
\label{subsubsec:Numerical}

\key{ADPARBASIS} 0 = \tv{ILIBAS}

\keyspace

\noindent In version {(v\codeversion)} of the code \pr{hfodd},
          the names of variables defining the oscillator frequency
          $\hbar\omega_0$, which are read under keyword
          \tk{SURFAC\_PAR}, have changed and now read \tv{INBASI},
          \tv{IZBASI}, and \tv{R0PARM}, see
          Section~II-3.5~\cite{(Dob97c)}. This allows for dynamically
          linking the original variables \tv{INNUMB} and \tv{IZNUMB}
          to neutron and proton numbers \tv{IN\_FIX} and
          \tv{IZ\_FIX}, see Section~II-3.1~\cite{(Dob97c)}, depending
          on the value of variable \tv{ILIBAS}, which has the allowed
          values of 0, 1, 2, and 3. Namely,\begin{itemize} \item for
          \tv{ILIBAS}=1 or \tv{ILIBAS}=3, \tv{INNUMB} is set to
          \tv{IN\_FIX}; otherwise it is set to \tv{INBASI}, \item for
          \tv{ILIBAS}=2 or \tv{ILIBAS}=3, \tv{IZNUMB} is set to
          \tv{IZ\_FIX}; otherwise it is set to
          \tv{IZBASI}.\end{itemize} For example, for \tv{ILIBAS}=0,
          the code defines $\hbar\omega_0$ (as before) by using
          variables read under keyword \tk{SURFAC\_PAR}, whereas for
          \tv{ILIBAS}=3 it does it by using the neutron and proton
          numbers \tv{IN\_FIX} and \tv{IZ\_FIX}.

\key{NEW\_WIGNER} 0 = \tv{NEWWIG}

\keyspace

\noindent For \tv{NEWWIG}=0 or 1, version {(v\codeversion)} of the
          code \pr{hfodd} uses the old and new method to calculate
          the Wigner $d$ functions, respectively, see
          Section~\ref{subsec:Wigner}.

\key{EVENQPNUMB} 0 = \tv{NUQEVE}

\keyspace

\noindent For \tv{ITWOBA}=1 or \tv{IPRGCM}>0, \tv{NUQEVE}=1 enforces
          even numbers of single-particle and quasiparticle states,
          which is needed for the implementation of the Pfaffian
          method, see Section~\ref{subsubsec:Pfaffian}. In version
          {(v\codeversion)} of the code \pr{hfodd}, \tv{NUQEVE}=1
          still requires \tv{ISIMPY}=0.

\subsection{Output parameters}
\label{subsubsec:Output}

\key{ALLNILABS} 0 = \tv{INUNIL}

\keyspace

\noindent For \tv{INUNIL}>0 and \tv{IREVIE}>0, up to \tv{INUNIL}=99
          non-dominant Nilsson labels are printed for
          each single-particle state on the
          \tk{REVIEWFILE}, see Section~II-3.9~\cite{(Dob97c)}.

\key{PRINT\_ELEC} 0 = \tv{IELPRI}

\keyspace

\noindent Switch \tv{IELPRI} defines the type of matrix elements of
          the electric, magnetic, surface, or Schiff operators
          between angular-momentum-projected states, which are
          printed in version {(v\codeversion)} of the code
          \pr{hfodd}: \begin{itemize} \item For \tv{IELPRI}=0, the
          code prints reduced matrix elements of the operators
          defined in the code, see~Section~IV-2.4~\cite{(Dob04d)}.
          \item For \tv{IELPRI}=1, the code prints reduced matrix
          elements of the standard operators that is, the particular
          units defined in Section~IV-2.4~\cite{(Dob04d)} are
          removed. \item For \tv{IELPRI}=2, the code prints the
          standard reduced transition rates
          BE$\lambda(I_i\longrightarrow{}I_f)$ and/or
          BM$\lambda(I_i\longrightarrow{}I_f)$. \item For
          \tv{IELPRI}=3, the code prints the standard spectroscopic
          matrix elements. \end{itemize}

\key{PRINTMATEL} 1, 0 = \tv{ILIMAM},\tv{IALLAM}

\keyspace

\noindent For \tv{ILIMAM}=0(1), the code does not (does) calculate
          reduced kernels and/or matrix elements that are not
          required for printing, as defined by the ranges of angular
          momenta specified in variables \tv{ISLPRI}, \tv{ISUPRI},
          see Section~VI-3.6~\cite{(Dob09g)}. For \tv{IALLAM}=0(1),
          the code does not (does) print reduced kernels and/or
          matrix elements between the "left" (bra) angular momentum
          larger than the "right" (ket) angular momentum. Those
          between smaller or equal angular momenta are always
          printed.

\key{PRINTALLRM} 0 = \tv{IPRALL}

\keyspace

\noindent For \tv{IPRALL}=1, the code prints all reduced kernels
          and/or matrix elements irrespective of restrictions
          otherwise imposed by switches \tv{IPRGCM}, \tv{IELPRI}, or
          \tv{IAXIAL}.

\key{REDMATSAVE} 0 = \tv{IWRIRM}

\keyspace

\noindent For \tv{IWRIRM}=1, an ASCII file with the reduced kernels
          and/or matrix elements, see keyword \tk{REDMATFILE}, is
          saved on disc after the angular-momentum projection is
          performed. \tv{ILIMAM}=1 allows for saving the reduced
          kernels and/or matrix elements that are not required for
          printing, see keyword \tk{PRINTMATEL}. \tk{IWRIRM}=1
          requires \tk{IPRGCM}>0 and is incompatible with
          \tk{IFTEMP}=1.

\key{REDMATFILE} HFODD.RED = \tv{FILRED}

\keyspace

\noindent CHARACTER*68 file name of the ASCII with the reduced
          kernels and/or matrix elements. Must start at the 13-th
          column of the data line.

\key{EFF\_G\_FACT} 0, 1., 1., 1. = \tv{IGYROS},\tv{GYRORP},\tv{GYRSPN},\tv{GYRSPP}

\keyspace

\noindent For \tv{IGYROS}=1, the standard single-particle orbital and
          spin gyroscopic factors, $g_\ell^{\nu,{\text{s.p.}}}$ and
          $g_s^{\nu,{\text{s.p.}}}$, respecively, for neutrons and
          protons, $\nu=n,p$, which are used to calculate the magnetic
          moments, see Section~IV-2.4~\cite{(Dob04d)}, are multiplied
          by the corresponding effective gyroscopic factors and read:
          \begin{equation}\begin{array}{rcl}
          g_\ell^p&=&g_\ell^{p,{\text{s.p.}}}*g_\ell^{p,{\text{eff}}}= + 1.000*\tv{GYRORP}, \\
          g_s^n&=&g_s^{n,{\text{s.p.}}}*g_s^{n,{\text{eff}}}= - 3.826*\tv{GYRSPN}, \\
          g_s^p&=&g_s^{p,{\text{s.p.}}}*g_s^{p,{\text{eff}}}= + 5.586*\tv{GYRSPP}.
          \end{array}\end{equation}

\key{QUASIPSAVE} $-$1 = \tv{IWRIQU}

\keyspace

\noindent For \tv{IWRIQU}=1, a binary quasiparticle file, see
          keyword \tk{QUASIPFILE}, is saved on disc after each
          iteration is completed. The file contains quasiparticle
          wave functions. For \tv{IWRIQU}=0, the file is saved only
          once, after all iterations are completed. For
          \tv{IWRIBA}=$-$1, the file is never saved. \tk{IWRIQU}=0 or
          1 requires \tk{IPAHFB}>0 and is incompatible with
          \tk{IFTEMP}=1 or \tk{IF\_RPA}=1. In version
          {(v\codeversion)} of the code \pr{hfodd}, \tv{IWRIQU}=0 or
          1 still requires \tv{ISIMPY}=1 and \tv{IPNMIX}=0.

\key{QUASIPFILE} HFODD.QUA = \tv{FILQUA}

\keyspace

\noindent CHARACTER*68 file name of the binary file that contains
          quasiparticle wave functions. Must start at the 13-th
          column of the data line.

\subsection{Starting, performing, stopping, and restarting iterations}
\label{subsubsec:Starting}

\key{MAXANTICON} 0.,  0 =  \tv{EPSCON},\tv{NUCONS}

\keyspace

\noindent For \tv{NUCONS}>0, iterations stop when the changes of the
          stability energy, Eq.~(I-37)~\cite{(Dob97b)}, stay below
          \tv{EPSCON}*\tv{EPSITE} for \tv{NUCONS} consecutive
          iterations. This option aims to stop iterations when the
          convergence is extremely slow or when the wave function
          infinitely alternates between two solutions both having the
          same value of the stability energy. Note that the
          alternating signs of the stability energy are recognised
          by the ping-pong divergence condition, see Sections~III-2.6
          and~III-3.1~\cite{(Dob00d)} and keyword \tk{PING\_PONG}.
          \tv{NUCONS}>0 requires \tv{EPSCON}>0.

\key{QUASISTABI} 0 = \tv{IQPSTA}

\keyspace

\noindent For \tv{IQPSTA}=1, expression (I-37)~\cite{(Dob97b)} for
          the stability energy is replaced by expression (\ref{eq:magic}),
          which is suitable for the HFB calculations.
          \tv{IQPSTA}=1 requires \tv{IPAHFB}>0. In version
          {(v\codeversion)} of the code \pr{hfodd}, \tv{IQPSTA}=1
          still requires \tv{IPNMIX}=0.

\key{SLOWALLFIL} 0.5,  0 = \tv{SLOWAL},\tv{I\_SLOW}

\keyspace

\noindent For \tv{I\_SLOW}=1, the rate of convergence is slowed down
          by a factor of \tv{SLOWAL} by mixing the mean field and
          pairing matrices on the harmonic-oscillator basis instead
          of mixing the mean-field potentials on the Gauss-Hermite
          spatial nodes, see keywords \tk{SLOW\_DOWN},
          Section~II-3.5~\cite{(Dob97c)}, \tk{SLOWLIPKIN},
          Section~VI-3.2~\cite{(Dob09g)}, \tk{SLOW\_PAIR},
          Section~IV-3.4~\cite{(Dob04d)}, and \tk{SLOWLIPMTD},
          Section~VIII-3.1.2~\cite{(Sch17d)}. \tv{I\_SLOW}=1 is
          incompatible with \tv{IBROYD}=1.

\key{BASIS\_SAVE} $-$1 = \tv{IWRIBA}

\keyspace

\noindent For \tv{IWRIBA}=1, the basis file, is saved on disc after
          each iteration is completed. The file contains Bohr
          deformation parameters that can be used to restart
          calculations with the basis deformation parameters equal to
          those read from the basis file, see Section~\ref{sec:basis}
          and keyword \tk{CONT\_BASIS}. For \tv{IWRIBA}=0, the file
          is saved only once, after all iterations are completed. For
          \tv{IWRIBA}=$-$1, the file is never saved.

\key{REPBASFILE} HFODD.BAP = \tv{FILBAP}

\keyspace

\noindent CHARACTER*68 file name of the basis file. Must start at the
          13-th column of the data line. For \tv{IBCONT}=1, see
          keyword \tk{CONT\_BASIS}, an ASCII basis file with the
          name defined in \tv{FILBAP} must exist, and will be read.
          If the filenames \tv{FILBAP} and \tv{FILBAC} are identical,
          the basis file will be subsequently overwritten as a new
          basis file.

\key{RECBASFILE} HFODD.BAC = \tv{FILBAC}

\keyspace

\noindent CHARACTER*68 file name of the basis file. Must start at the
          13-th column of the data line. For \tv{IWRIBA}=0 or 1, an
          ASCII basis file is saved, see keyword \tk{BASIS\_SAVE}.

\key{CONT\_BASIS} 0 = \tv{IBCONT}

\keyspace

\noindent For \tv{IBCONT}=1, the Bohr deformation parameters stored
          on the basis file are used to restart calculations with the
          basis deformation parameters equal to those read from the
          basis file, see Section~\ref{sec:basis}. For \tv{IBCONT}=1,
          an ASCII basis file with the name defined in \tv{FILBAP}
          must exist, and it will be read. \tv{IBCONT}=1 is
          incompatible with \tv{ICONTI}=0.

\subsection{Miscellaneous}
\label{subsubsec:Miscellaneous}

\key{ANTISYMPAI}  0  =  \tv{KAPASY}

\keyspace

\noindent For \tv{KAPASY}=1, the pairing tensor is antisymmetrised,
          which removes its possible nonzero symmetric component
          that can appear for the quasiparticle cutoff,
          cf.~Section~IV-3.1~\cite{(Dob04d)} and
          Refs.~\cite{(Bor05),(Dob13)}.

\key{FERMICUT}  0  =  \tv{LAMCUT}

\keyspace

\noindent For \tv{LAMCUT}=1, the quasiparticle cutoff,
          see~Section~IV-3.1~\cite{(Dob04d)}, is applied relatively
          to the proton and neutron Fermi energies, and not relatively
          to the zero of the equivalent single-particle spectrum,
          which is the default.

\key{LAN4SCALED}   0  =  \tv{LANSCA}

\keyspace

\noindent For \tv{LANSCA}=1, the Landau parameters,
          see~Sections~IV-2.8 and IV-3.1~\cite{(Dob04d)}, are used
          for scaled, see~Section~II-3.2~\cite{(Dob97c)}, coupling
          constants of the functional and not for the unscaled ones,
          which is the default.

\key{NOZEROPAIR}    0  =  \tv{NOZEPA}

\keyspace

\noindent For \tv{NOZEPA}=1, the zero-range pairing force is
          neglected regardless of the values of pairing strengths
          defined in Section~IV-3.1~\cite{(Dob04d)}.

\subsection{New features of previously implemented keywords}
\label{subsubsec:Previous}

\key{OPTI\_GAUSS} 1 = \tv{IOPTGS}

\keyspace

\noindent Apart from the value of \tv{IOPTGS}=1, implemented in
          Section~II-3.5~\cite{(Dob97c)}, whereupon expression (I-94)
          was used to calculate the orders of the Gauss-Hermite
          integrations in the three Cartesian directions, that is,
          \tv{NXHERM}=2*\tv{NXMAXX}+2,
          \tv{NYHERM}=2*\tv{NYMAXX}+2,
          \tv{NZHERM}=2*\tv{NZMAXX}+2, the version
          {(v\codeversion)} of the code \pr{hfodd} now accepts a new
          value of \tv{IOPTGS}=2, for which
          \tv{NXHERM}=3*\tv{NXMAXX}+2,
          \tv{NYHERM}=3*\tv{NYMAXX}+2,
          \tv{NZHERM}=3*\tv{NZMAXX}+2. The former (latter) values ensure exact
          Gauss-Hermite integrations of the two-body (three-body) zero-range terms
          with second-order gradients. The latter values are thus suitable for calculations
          described in Section~\ref{subsec:gradient}. For \tv{IOPTGS}=0 or \tv{IREAWS}=1,
          expressions above do not overwrite values read under keyword \tk{GAUSHERMIT}.

\key{MULTCONSTR}  2, 0, 0.01, 42.0, 1  =
                  \tv{LAMBDA}, \tv{MIU},
                  \tv{STIFFQ}, \tv{QASKED}, \tv{IFLAGQ}

\keyspace

\noindent Apart from the value of \tv{IFLAGQ}=1, previously
          implemented in Section~II-3.7~\cite{(Dob97c)}, the version
          {(v\codeversion)} of the code \pr{hfodd} now accepts a new
          value of \tv{IFLAGQ}=$-1$ mentioned in
          Section~\ref{sec:basis}, whereupon the value of \tv{QASKED}
          for  $\lambda$=\tv{LAMBDA} and $\mu$=\tv{MIU} is used for a
          definition of the HO basis, whereas the corresponding
          constraint is ignored. For \tv{IFLAGQ}=0, the value of
          \tv{QASKED} is ignored. Recall that unless \tv{IFLAGQ}=0 is
          explicitly set in the input data file for \tv{LAMBDA}=2 and
          \tv{MIU}=0, the constraint on $Q_{20}$=42\,b would be
          active by default.

\key{FREQBASIS} 1.0, 1.0, 1.0, 0 = \tv{BASINX}, \tv{BASINY}, \tv{BASINZ}, \tv{INPOME}

\keyspace

\noindent Apart from the value of \tv{INPOME}=1, previously
          implemented in Section~VIII-3.1.4~\cite{(Sch17d)},
          and the default value of \tv{INPOME}=0,
          the version {(v\codeversion)} of the code \pr{hfodd}
          now accepts new values of \tv{INPOME}=2,\ldots,7
          described in Section~\ref{sec:basis}.

\key{REVIEW}  2 = \tv{IREVIE}

\keyspace

\noindent Apart from the values of \tv{IREVIE}=0, 1, and 2, previously
          implemented in Section~II-3.9~\cite{(Dob97c)},
          the version {(v\codeversion)} of the code \pr{hfodd}
          now accepts new values of \tv{IREVIE}=$-2$, and 3,\ldots,8,
          which allow for printing on the REVIEW file the following
          {\em additional} information:
          \begin{itemize}
          \item For \tv{IREVIE} = $-2$, the quasiparticle data are printed.
          \item For \tv{IREVIE}$\geq$3, the $x$ and $z$ single-particle alignment data are printed.
          \item For \tv{IREVIE}$\geq$4, the proton-neutron single-particle data are printed.
          \item For \tv{IREVIE}$\geq$5, the proton-neutron single-particle alignment data are printed.
          \item For \tv{IREVIE}$\geq$6, the integration points are printed.
          \item For \tv{IREVIE}$\geq$7, the integration weights are printed.
          \item For \tv{IREVIE} = 8,    the densities are printed.
          \end{itemize}

\key{BASIS\_SIZE} 15, 301, 800.0 = \tv{NOSCIL}, \tv{NLIMIT}, \tv{ENECUT}

\keyspace

\noindent Apart from the values of \tv{NLIMIT}>0, previously
          implemented in Section~II-3.6~\cite{(Dob97c)},
          the version {(v\codeversion)} of the code \pr{hfodd}
          now accepts the value of \tv{NLIMIT}=0, whereupon
          the value of \tv{NLIMIT} is instantly recalculated
          to \tv{NLIMIT}=((\tv{NOSCIL}+1)*(\tv{NOSCIL}+2)*(\tv{NOSCIL}+3))/6,
          which corresponds to the number of states of a spherical HO
          with the total number of quanta not exceeding \tv{NOSCIL}.

\key{CONTLIPKIN} 0 = {\tv{ILCONT}}

\keyspace

\noindent Apart from the values of \tv{ILCONT}=0 or 1, previously
          implemented in Section~VI-3.8~\cite{(Dob09g)}, the version
          {(v\codeversion)} of the code \pr{hfodd} now accepts the
          value of \tv{ILCONT}=2. This value does not request reading
          the LIPKIN FILE, but allows for reading the Lipkin-Nogami
          parameters $\lambda_2$ from the RECORD FILE stored in a
          previous run. In conjunction with reading the FIELDS FILE
          (\tv{IFCONT}=1), \tv{ILCONT}=2 thus allows for a smooth
          continuation of the Lipkin-Nogami calculations.
          \tv{ILCONT}>0 is incompatible with either of
          \tv{LIPKIN}=\tv{LIPKIP}=0 or \tv{IPCONT}=0.

\key{HFB}  0  = \tv{IPAHFB}

\keyspace

\noindent Apart from the values of \tv{IPAHFB}=0 or 1, previously
          implemented in Section~IV-3.2~\cite{(Dob04d)}, the version
          {(v\codeversion)} of the code \pr{hfodd} now accepts the
          value of \tv{IPAHFB}=2, whereupon the HFB densities are
          summed up in the canonical basis. \tv{IPAHFB}>0 requires
          \tv{IPAIRI}=1, see Section~II-3.3~\cite{(Dob97c)}, and
          \tv{IPAHFB}=2 is incompatible with \tv{ITWOBA}=1,
          \tv{IMFHFB}=1, \tv{IFSHEL}>0, or \tv{IPNMIX}=1. However,
          the HFB method was already implemented in all symmetries
          and thus in the version {(v\codeversion)} of the code
          \pr{hfodd} the restriction to \tv{ISIMPY}=1, specified in
          Section~IV-3.2~\cite{(Dob04d)}, was lifted.

\key{PROJECTGCM}
\nopagebreak
\noindent\begin{tabular}{@{}lll}
             & 0, 0, 0, ~~ 1, 1, 0, ~~ 1, 1, 0 &  \\
             & {\tv{IPRROT}}, {\tv{IPROMI}}, {\tv{IPROMA}},
             & {\tv{NUAKNO}}, {\tv{NUBKNO}}, {\tv{KPROJE}},                 \\
            && {\tv{IFRWAV}}, {\tv{ITOWAV}}, {\tv{IWRWAV}}
\end{tabular}
\keyspace

\noindent The name of the first variable was changed to \tv{IPRROT}
          and thus this variable was made independent form the first
          variable read under keyword \tk{PROJECTGCM}, see
          Section~\ref{subsubsec:ISOSTZ}. For \tv{IPRROT}=0,
          the remaining input data read under keyword \tk{PROJECTGCM}
          are now ignored.

          For {\tv{IPRGCM}}=2, apart from a positive value of
          \tv{IFRWAV}, previously implemented in
          Section~VI-3.2~\cite{(Dob09g)}, the version
          {(v\codeversion)} of the code \pr{hfodd} now accepts its
          negative value, whereupon the calculation of the GCM
          kernels is performed between states with labels
          $-$\tv{IFRWAV} and \tv{ITOWAV} only, and not between {\em
          all} states with labels from a postive label \tv{IFRWAV} to
          \tv{ITOWAV}. In the version {(v\codeversion)} of the code
          \pr{hfodd}, the angular-momentum projection (AMP) was
          implemented for the GCM kernels, so for
          {\tv{NUAKNO}}$\neq$1 or {\tv{NUBKNO}}$\neq$1,
          {\tv{IPRGCM}}=2 is now allowed.

\key{PROJECTISO} 0, 2, 1, 1.E-6, 0, 0 =
                 \tv{IPRISO}, \tv{ISOSAD}, \tv{NBTKNO}, \tv{EPSISO}, \tv{ICSKIP}, \tv{IFERME}

\keyspace

\noindent The name of the first variable was changed to \tv{IPRISO}
          and thus this variable was made independent form the first
          variable read under keyword \tk{PROJECTGCM}, see
          Section~\ref{subsubsec:ISOSTZ}. For \tv{IPRISO}=0,
          the remaining input data read under keyword \tk{PROJECTISO}
          are now ignored.

\key{SAVEKERNEL} 0 = {\tv{ISAKER}}

\keyspace

\noindent Apart from the values of \tv{ISAKER}=0 or 1, previously
          implemented in Section~VI-3.2~\cite{(Dob09g)}, the version
          {(v\codeversion)} of the code \pr{hfodd} now accepts the
          value of \tv{ISAKER}=2, whereupon the \ti{kernel file} is
          stored on the disc in a new format. Although the old
          format, requested by \tv{ISAKER}=1, is still supported, it
          should be considered obsolete. The use of the old format is
          not recommended because some newly developed features may
          then be improperly stored. In particular, \tv{ISAKER}=1 is
          incompatible with \tv{KETAJ2}=1, \tv{KETAT2}=1,
          \tv{NPNKNO}>1, \tv{NTZKNO}>1, or \tv{NPAKNO}>1.

          For {\tv{ISAKER}}=2 and {\tv{IPAKER}}=1 (see Section~VI-3.2~\cite{(Dob09g)}),
          for all values of indices "t" the code attempts reading the \ti{kernel file}s
          Nxxxxxt-Lyyy-Rzzz-//{\tv{FILKER}}, where // denotes concatenated
          strings. The one-, three-, or five-digit indices are:
          \begin{itemize}
          \item "xxxxx" is the consecutive index of the \ti{kernel file}, which  is
                equal to \tv{KFIKER} (see Section~VII-3.2~\cite{(Sch12c)}),
          \item "t" is the number from 0 to 9 of the consecutive
                     file having the given index "xxxxx",
          \item "yyy" is the number  of  the  left  wave  function,
          \item "zzz" is the number of  the  right  wave  function.
          \end{itemize}
          In the work directory, the file names for all indices "xxxxxt"
          are scanned, starting from xxxxx0. The kernels stored in these
          files are read into memory and are not recalculated. Those
          that have not been found in the \ti{kernel file}s
          are calculated and stored in the kernel file with the
          lowest available index "t". In this way, one can submit
          many parallel jobs, see the keyword {\tk{PARAKERNEL}} (see
          Section~VI-3.2~\cite{(Dob09g)}). The results are then
          collected in different \ti{kernel file}s with indices "t"
          attributed automatically. If any of the jobs is terminated
          before completing its task, the same input data can be
          resubmitted and the calculation automatically continues
          from the point where it has been interrupted. Once all the
          kernels will have been calculated (with {\tv{IPAKER}}=1),
          which requires a large CPU time, the AMP can be performed
          (with {\tv{IPAKER}}=0) within a very small CPU time by
          reading, again automatically, all the created \ti{kernel
          file}s with indices "xxxxxt". At the AMP stage, \tv{KFIKER}
          denotes the maximum index "xxxxx" of the \ti{kernel file}s
          that were stored. Note that if at the AMP stage any kernels
          are missing, the code will attempt to calculate them. For
          {\tv{ISAKER}}=2, this feature can be overridden by using
          {\tv{IPAKER}}=$-$1) instead of {\tv{IPAKER}}=0), whereupon if
          any kernels are missing the code will stop. {\tv{ISAKER}}=2
          requires {\tv{IPRGCM}}$>$0 and 0$<${\tv{KFIKER}}$<$99999.

\key{PARAKERNEL} 0, 1, 1, 1, 1 = {\tv{IPAKER}}, {\tv{NUASTA}}, {\tv{NUASTO}},
                                                {\tv{NUGSTA}}, {\tv{NUGSTO}}
\keyspace

\noindent Apart from the values of \tv{IPAKER}=0 or 1, previously
          implemented in Section~VI-3.2~\cite{(Dob09g)}, the version
          {(v\codeversion)} of the code \pr{hfodd} now accepts the
          value of \tv{IPAKER}=$-$1.  For {\tv{ISAKER}}=2,
          \tv{IPAKER}=$-$1 is equivalent to \tv{IPAKER}=0, however,
          if in the earlier parallel runs (with {\tv{IPAKER}}=1) any
          kernels were not yet calculated, the code will stop instead
          of attempting to calculate them.

\section{Fortran Source Files}
\label{sec:source_files}

The \ti{FORTRAN source} of version {(v\codeversion)} of the code \pr{hfodd}
is provided in the file \tf{hf{\execversion}.f}, and its accompanying modules are:
\begin{itemize}
\item \tf{hfodd\_sizes\_{\sizeversion}.f90}: Static array size
declarations. Contains all {\tt PARAMETER} statements controlling the
sizes of all statically allocated arrays (and some of the dynamically
allocated arrays) used in the code. The static allocations are
maintained in the code because the compiler optimisation options are
often more efficient when the dimensions of arrays are known to the
compiler. In practice, only a few size declarations need to be defined by
the user. In module \tf{hfodd\_sizes\_{\sizeversion}.f90}, such declarations
are collected at the beginning of the module and read:
\newline\begin{tabular}{l@{ }l}
\tv{NDMAIN}, & Maximum number of the HO shells,                                              \\
\tv{NDBASE}, & Maximum number of the HO basis states,                                        \\
\tv{NDSTAT}, & Maximum number of the HF states without spin,                                 \\
\tv{NDXHRM}, & Maximum number of the Gauss-Hermite nodes in the $x$ direction,               \\
\tv{NDYHRM}, & Maximum number of the Gauss-Hermite nodes in the $y$ direction,               \\
\tv{NDZHRM}, & Maximum number of the Gauss-Hermite nodes in the $z$ direction,               \\
\tv{NDPROI}, & Maximum doubled spin in the angular-momentum projection (AMP),                \\
\tv{NDAKNO}, & Maximum number of the nodes in the $\alpha$ and $\gamma$ Euler AMP angles,    \\
\tv{NDBKNO}, & Maximum number of the nodes in the $\beta$ Euler AMP angle,                   \\
\tv{NDPROT}, & Maximum doubled isospin in the isospin projection (IP),                       \\
\tv{NDATKN}, & Maximum number of the nodes in the $\alpha_T$ and $\gamma_T$ Euler IP angles, \\
\tv{NDBTKN}, & Maximum number of the nodes in the $\beta_T$ Euler IP angles.
\end{tabular}\newline
Any version 5 of the module, \tf{hfodd\_sizes\_5.f90}, can be upgraded to version 7
by copying a few lines of code located at the end of module \tf{hfodd\_sizes\_{\sizeversion}.f90}.

The code \pr{hfodd} can be perfectly well run with array sizes
smaller than the maximum ones specified above; the only consequence
would be a non-optimal memory usage. If a given requested array size
exceeds the maximum, the code stops and prints the new maximum size
that has to be used at compilation.

\item \tf{hfodd\_modules\_{\modulesversion}.f}: Definitions of memory-consuming
modules. Defines, among others, the matrices of the Bogolyubov transformation,
the eigenvectors of the HF and HFB equations, etc.

\item \tf{hfodd\_hfbtho\_{\hfbthoversion}.f90}: \pr{hfbtho} DFT solver based
on version 200d published in Ref.~\cite{(Sto12)}.

\item \tf{hfodd\_interface\_{\interfaceversion}.f90}: Interface between the
\pr{hfbtho} and \pr{hfodd} solvers. Contains the routine to transform the HFB
matrix from the HO basis used in \pr{hfbtho} (\pr{hfodd}) to the basis used
in \pr{hfodd} (\pr{hfbtho}).

\item \tf{hfodd\_functional\_{\functionalversion}.f90}: Interface to UNEDF
functionals.

\item \tf{hfodd\_mpiio\_{\mpiioversion}.f90}: IO interface in MPI calculations.
Contains the routine to read input data for parallel \pr{hfodd} calculations.

\item \tf{hfodd\_mpimanager\_{\mpimanagerversion}.f90}: MPI toolkit. Defines
the list of MPI tasks based on the data read in the parallel input file
\tf{hfodd\_mpiio.d}.

\item \tf{hfodd\_shell\_{\shellversion}.f}: Toolkit for the shell correction.

\item \tf{hfodd\_SLsiz\_{\SLsizversion}.f}: Toolkit to incorporate ScaLAPACK capabilities.

\item \tf{hfodd\_fission\_{\fissionversion}.f90}: Toolkit for fission
calculations. Contains several routines to compute fission fragments properties
such as charge, mass, total energy, interaction energy; the routines needed to
use a constraint on the number of particles in the neck; the routines used for
the quantum localization method.

\item \tf{hfodd\_pairs\_{\pairsversion}.f90}: Toolkit for defining
various derived types related to pairs, lists of pairs, and lists
of lists of pairs, as well as the routines needed to manipulate these
objects.

\item \tf{hfodd\_pnp\_{\pnpversion}.f90}: Toolkit for particle number projection
(not supported in the present version {(v\codeversion)} of the
code \pr{hfodd}).

\item \tf{hfodd\_fits\_{\fitsversion}.f90}: Fit module. Allows the code
\pr{hfodd} to work as a routine in an external program.

\item \tf{hfodd\_lipcorr\_{\lipkinversion}.f90}: Toolkit for the Lipkin method
and Pfaffian overlap calculations.

\item \tf{hfodd\_tgrad\_{\tgradversion}.f90}: Toolkit for the
three-body gradient terms.

\item \tf{hfodd\_wigner\_{\wignerversion}.f90}: Toolkit for the
Wigner functions.

\end{itemize}

The \ti{FORTRAN source} of version {(v\codeversion)} of the code
\pr{hfodd} contains numerous undocumented and untested features that
are under development. The user should not attempt to activate or
reverse-engineer these features, because this can certainly lead
to an unpredictable behavior of the code and even damage to
computer hard drive.

\section{Acknowledgments}
\label{sec:acknowledgments}

\bigskip
We would like to thank Nicolas Schunck for performing benchmark
calculations with his codes.
This work was partially supported by the STFC Grant Nos.~ST/M006433/1
and~ST/P003885/1, and by the Polish National Science Centre under
Contract No.~2018/31/B/ST2/02220.
Work of MB and KB was supported by the {\it Agence Nationale de la Recherche}
under grant No.~19-CE31-0015-01 (NEWFUN).
We acknowledge the CSC-IT Center for
Science Ltd., Finland, for the allocation of computational resources.
We gratefully acknowledge support from the CNRS/IN2P3 Computing Center
(Lyon - France) for providing computing and data-processing resources needed for this work.
This project was partly undertaken on the Viking Cluster, which is a high
performance compute facility provided by the University of York. We
are grateful for computational support from the University of York
High Performance Computing service, Viking and the Research Computing
team.

\section*{References}

\bibliographystyle{iopart-num}
\providecommand{\newblock}{}

\end{document}